\input{epsf}

\documentclass[12pt,preprint]{aastex}

\shorttitle{Fast Rotator $\alpha$ Leo}
\shortauthors{McAlister et al.}

\received{}
\accepted{}

\begin{document}

\title{First Results from the CHARA Array. I. An Interferometric and 
Spectroscopic Study of the Fast Rotator $\alpha$ Leonis (Regulus)}

\author{H. A. McAlister, T. A. ten Brummelaar, D. R. Gies\altaffilmark{1}, 
W. Huang\altaffilmark{1}, W. G. Bagnuolo, Jr., \\ M. A. Shure,
J. Sturmann, L. Sturmann, N. H. Turner, S. F. Taylor, D. H. Berger, \\
E. K. Baines, E. Grundstrom\altaffilmark{1}, C. Ogden}
\affil{Center for High Angular Resolution Astronomy, 
Georgia State University, \\
P.O. Box 3969, Atlanta, GA 30302-3969}
\email{hal@chara.gsu.edu, theo@chara-array.org, 
gies@chara.gsu.edu, huang@chara.gsu.edu, bagnuolo@chara.gsu.edu, mashure@yahoo.com,
judit@chara-array.org, sturmann@chara-array.org, nils@chara-array.org, taylor@chara.gsu.edu, 
berger@chara-array.org, baines@chara.gsu.edu, erika@chara.gsu.edu, ogden@chara.gsu.edu}

\author{S. T. Ridgway}
\affil{Kitt Peak National Observatory, National Optical Astronomy Observatory, \\
P.O. Box 26732, Tucson, AZ 85726-6732}
\email{sridgway@noao.edu}

\and

\author{G. van Belle}
\affil {Michelson Science Center, California Institute of Technology, \\
770 S. Wilson Ave, MS 100-22, Pasadena, CA 91125}
\email {gerard@ipac.caltech.edu}

\altaffiltext{1}{Visiting Astronomer, Kitt Peak National Observatory, 
National Optical Astronomy Observatory, operated by the Association of 
Universities for Research in Astronomy, Inc., under contract with the National Science Foundation.}

\newpage

\begin{abstract}
We report on $K$-band interferometric observations of the bright, 
rapidly rotating star Regulus (type B7~V) made with the CHARA Array on Mount Wilson, California. 
Through a combination of interferometric and spectroscopic measurements, we have 
determined for Regulus the equatorial and polar diameters and temperatures, the 
rotational velocity and period, the inclination and position angle of the spin axis, 
and the gravity darkening coefficient. These first results from the CHARA Array provide 
the first interferometric measurement of gravity darkening in a rapidly rotating star 
and represent the first detection of gravity darkening in a star that is not a member 
of an eclipsing binary system.
\end{abstract}


\setcounter{footnote}{1}
\section{Introduction} 

\subsection{The CHARA Array}

Georgia State University's Center for High Angular Resolution Astronomy (CHARA) operates 
an optical/IR interferometric array on the grounds of Mount Wilson Observatory in the 
San Gabriel Mountains of southern California. The six light-collecting telescopes of 
the CHARA Array, each of 1-m aperture, are distributed in a Y-shaped configuration 
providing 15 baselines ranging from 34.1 to 330.7~m. Three of these baselines formed by 
the outer telescopes of the Y are in excess of 300~m. Ground breaking for the facility 
occurred in 1996 July, and the sixth and final telescope became fully integrated into the 
Array in 2003 December, thus signaling completion of construction and readiness for 
on-going science observations. Numerous technical reports and updates have 
been published throughout the design and construction phases of this project, the most recent of 
which is by McAlister et al.\ (2004), with many internal project reports available at 
CHARA's website\footnote{http://www.chara.gsu.edu/CHARA/techreport.html}. 
A companion paper \citep{bru04} describes the full complement of technical 
and performance details of the instrument.

With its long baselines, currently the longest operational $K$-band baselines in the world, and 
its moderate apertures, the CHARA Array is well positioned for observations of main sequence 
stars not previously accessible to such high-resolution scrutiny. CHARA inaugurated its first 
diverse observing season in the spring of 2004 with a fully scheduled semester of science observations 
targeting stellar diameters, young stellar objects, and rapidly rotating stars.  That semester was 
immediately preceded by an observing campaign directed entirely at Regulus, a very rapidly rotating star 
expected to exhibit marked rotational oblateness and a resulting surface flux variation associated with 
gravity darkening. This paper is a report of the outcome of those observations of Regulus and 
represents the first scientific results from the CHARA Array.

\subsection{Regulus}

The star Regulus ($\alpha$ Leo, HR 3982, HD 87901) is of spectral type B7~V (Johnson \& Morgan 1953) 
or B8~IVn \citep{gra03}, and is a well-known rapid rotator. 
The difficulty of determining the projected rotational velocity $V\sin i$ is exemplified by the wide range of values 
published for this star, with velocities in the literature ranging from 
$249\pm9$ km~s$^{-1}$ (Stoeckley, Carroll, \& Miller 1984)
to $350\pm25$ km~s$^{-1}$ \citep{sle63}. The first attempts to model realistically 
a line profile derived from a rotating star were by Elvey (1930) who employed the 
method proposed by Shajn \& Struve (1929) based upon dividing a photosphere into strips 
whose Doppler shifted contributions are weighted by the relative areas of these strips. 
Slettebak (1949) very significantly extended this analysis for rapidly rotating O and B stars 
to include the effects of limb darkening, gravity darkening based on Roche models incorporating 
the von Zeipel Effect \citep{zei24}, and differential rotation. Slettebak (1966) subsequently 
showed that the envelope of the most rapidly rotating stars of a given spectral type most 
closely approached the threshold of breakup velocities for stars of early- to mid-B type. 
Regulus clearly falls in the domain of stars rotating close to breakup speed and can thus 
be expected to show significant rotationally induced oblateness and gravity darkening. 

The Washington Double Star Catalog\footnote{http://ad.usno.navy.mil/wds/wds.html; 
maintained at the U.\ S.\ Naval Observatory} lists three companions to Regulus, none of 
which has exhibited significant orbital motion since its first detection. The B component, 
$\sim175$~arcsec from Regulus, is HD~87884, a K2~V star whose proper motion, radial velocity 
and spectroscopic parallax are consistent with its being a physical rather than optical companion. 
On the other hand a comparison of ages of components A and B has found the discordant values 
of 150 and 50 Myr, respectively \citep*{ger01}. Component B is accompanied by a faint companion 
($V=+13.1$) comprising the subsystem BC that closed from 4.0 arcsec to 2.5 arcsec during the 
period 1867 to 1943 and has apparently not been measured in the last 60 years. At the distance of 
Regulus, the apparent magnitude of component C implies a star of approximate spectral type M4~V. 
Only a single measurement exists for a D component, more than 200 arcsec from Regulus, whose 
existence as a physical member of the $\alpha$~Leo system might therefore be suspect.

The Bright Star Catalog (Hoffleit 1982) notes that Regulus is a spectroscopic binary. 
We have found no reference to this in the literature except a flag in the bibliographic 
catalog of Abt \& Biggs (1972) indicating the ``progressive change'' of radial velocity as 
reported by Maunder (1892).  After conversion to modern units (Maunder used German geographical 
miles per second), the measured velocities declined from $+40$~km~s$^{-1}$ to $-9$~km~s$^{-1}$ between 
1875 and 1890.  In light of the difficulty of accurately measuring velocities from hydrogen lines 
compounded by the very high rotational speed and the state of the art of the technique 
more than a century ago, the conclusion that those data indicate orbital motion seems unlikely. 
Furthermore, the General Catalogue of Stellar Radial Velocities (Wilson 1963) does not flag 
the star as a spectroscopic binary. On the other hand, Regulus may represent a case of neglect 
by radial velocity observers due to its brightness and extreme rotational broadening. 
If the velocities were to hint at a closer companion than the AB system, the corresponding 
orbital period would be on the order of a year with a separation in the range 50 to 70 mas, 
the regime of detectability by speckle interferometry. However, no closer companions have 
been reported from lunar occultation observations or from the CHARA speckle interferometry 
survey of bright stars (McAlister et al.\ 1993). In this angular separation regime, the fringe packets 
from Regulus and a very much fainter close companion would be non-overlapping in the fringe scan, 
and the companion's presence would have no effect on our analysis of stellar shape and 
surface brightness.  One of the goals of this study has been to determine the orientation of 
the spin axis of Regulus. Owing to the lack of sufficient motion within the Regulus 
multiple star system and the apparent lack of any unknown closer companions, it will 
not be possible to compare rotational and orbital angular momentum vectors.

Regulus possesses an intrinsic brightness and relative nearness to the Sun to make it 
an ideal candidate for diameter determination from long-baseline interferometry. 
Therefore, the star was among the classic sample of bright stars whose diameters were 
measured by Hanbury Brown and his colleagues at the Narrabri Intensity Interferometer 
nearly four decades ago (Hanbury Brown 1968; Hanbury Brown et al.\ 1967; 
Hanbury Brown, Davis, \& Allen 1974). In the first of these papers, the implications 
of a very high rotational velocity for the derived diameter were clearly recognized 
and discussed semi-quantitatively, but the resulting diameter of 1.32$\pm$0.06 mas, 
measured at $\lambda$ = 438.5 nm for a uniform disk (UD), was judged to incorporate 
insufficient baseline sampling to be interpreted as other than circularly symmetric. 
A value of 1.37$\pm$0.06 mas was determined for a limb-darkened disk (LD). Because gravity 
darkening brightens the poles and thus counteracts the star's oblateness, 
it was thought that the maximum change in apparent angular diameter within the primary lobe was only 6\%, 
which was judged to be marginal to detect in those observations.  Furthermore, measurements beyond the first lobe were 
impossible because of the limited sensitivity of the Narrabri instrument (in which the 
response varies as the square of the visibility).  

The other direct measurements of the angular diameter of Regulus come from lunar occultations.  
Although an early measurement \citep{ber70} yielded a diameter of 1.7$\pm$0.5 mas, the only two reliable 
measurements of Regulus have been made by Radick (1981) and Ridgway et al.\ (1982).  Radick (1981) 
obtained a diameter of 1.32$\pm$0.12 mas (UD) at $\lambda$ = 576.8 nm and 1.37$\pm$0.11 (UD) at 
$\lambda$ = 435.6 nm.  These translate to 1.36 and 1.43 mas (LD), respectively. The angle of local 
moon normal to the star was small, thus this observation sampled the minor axis of the star.  
A second set of observations obtained several months later at KPNO by Ridgway et al.\ (1982) 
with the 4-m telescope in $H$-band (1.67 $\mu$m) lead 
to a diameter of 1.44$\pm$0.09 mas (UD) or 1.46 (LD).  The angle of approach was somewhat steeper in 
this instance, about 32$^\circ$ (D.\ Dunham, 2004, private communication).   It is tempting to suppose that the 
somewhat higher value in this case is due to the geometry, but the observations obviously agree within stated errors.  
The simple average of all three observations (averaging the two from Radick) is 1.42 mas (LD), 
which agrees well with our diameter measurements discussed below.    

A detailed consideration of the response of a long-baseline interferometer to a 
rotationally distorted star was first undertaken by Johnston \& Wareing (1970) who 
identified Regulus and Altair ($\alpha$~Aql), along with the somewhat fainter star 
$\zeta$ Oph, as nearly ideal targets for an interferometer with sensitivity beyond 
that of the Narrabri instrument. They argued that the high projected rotational 
velocities of these stars implied inclinations of their rotational axes with respect 
to the line of sight of nearly 90$^\circ$.

Two rapidly rotating stars have now been observed with the new generation of interferometers 
envisioned by Johnston \& Wareing (1970). Van Belle et al.\ (2001) have used the Palomar 
Testbed Interferometer (PTI) to measure the oblateness of Altair and to derive a $V\sin i$ 
independent of spectroscopy on the basis of a rapidly rotating Roche model to which the
measured interferometric visibilities were fit. The PTI results led to the determination 
of the star's polar diameter of 3.04$\pm$0.07 mas with an equatorial diameter 14\% larger. 
The Very Large Telescope Interferometer (VLTI) has been used by Domiciano de Souza et al.\ (2003) 
to study the oblateness of the Be star Achernar ($\alpha$~Eri). A rotating Roche model with 
gravity darkening was applied to the collective VLTI visibilities following an initial 
simplified uniform disk diameter fit to individual visibilities with the conclusion that 
the Roche approximation does not pertain in the case of Achernar. Achernar was found to 
have a polar angular diameter of 1.62$\pm$0.06 mas with an extraordinary oblateness ratio of 1.56$\pm$0.05.

The smaller diameter of Regulus in comparison with Altair and Achernar calls for an interferometer 
with baselines longer than those possessed by the PTI and VLTI instruments. At the longest baselines 
of the CHARA Array, access is obtained in the $K$-band infrared to visibilities descending to the 
first null in the regime in which the departures from a uniform disk expected for a rapid rotator 
become increasingly more pronounced. With this in mind and inspired by the interest generated from 
the PTI and VLTI results, Regulus was selected as an ideal target for an extensive observing 
campaign with the CHARA Array. The combination of the resulting CHARA interferometric data with 
a number of constraints resulting from spectroscopy of Regulus have permitted us to determine 
the star's polar and equatorial angular diameters and temperatures, inclination angle of the 
spin axis with respect to the line of sight, position angle of the spin axis in the plane of the sky, 
equatorial rotational velocity, and gravity darkening index.

\section{Interferometric Observations} 

Regulus was observed on a total of 22 nights between 2004 March 10 and 2004 April 16 using 10 of 
the 15 baselines available from the CHARA array. A calibrator star, HD~83362 (G8~III, $V=+6.73$, $K=+4.60$), 
was selected on the basis of its predicted small angular diameter and its lack of known stellar companions. 
This star is located $7\fdg4$ from Regulus. No suitable calibrator was found closer in angular separation. 
A third object, HD~88547 (K0~III, $V=+5.78$, $K=+2.97$), was chosen as a check star on the basis of having a 
predicted angular diameter comparable to that of Regulus but not expected to show rotational oblateness due 
to its low $V\sin i = 2.5\pm0.8$ km~s$^{-1}$ \citep{hen00}. These three objects were observed in a 
sequence so as to provide a series of time-bracketed observations permitting the conversion from raw 
to calibrated visibility for each observation of Regulus and the check star.

All observations were obtained in the $K$-band infrared using a $K^{\prime}$-band filter whose 
effective wavelength was determined to be 2.1501 $\mu$m for a star of spectral type B7. 
This value is based upon the atmospheric transmission, the measured filter transmission, 
the detector DQE, and from data provided by the vendors for the various optical surfaces 
encountered along the light path of each arm of the interferometer. The stellar photon count 
rate is also included, and the increase in effective filter wavelength to 2.1505 $\mu$m for 
the late-type giants is considered to be negligible here.

Interference fringes were obtained using a classical two-beam interferometer configuration 
in which light is detected emerging from both sides of a beam splitter with a PICNIC HgCdTe 
256$\times$256 hybrid focal plane array (sensitive to the wavelength range 1.0 to 2.5 $\mu$m) developed by Rockwell Scientific 
Company \citep{stu02}. After open-loop path length difference compensation is attained with 
the optical delay lines, the zero path difference position is scanned with a dither mirror 
in order to produce fringes of a selectable frequency, with 150~Hz fringes being utilized 
for all the data in this analysis. A single dataset consists of some 200 fringe scans preceded 
and followed by shutter sequences to measure the light levels on each input side of the beam splitter. 
These sequences permit the calculation of a visibility adjustment factor to correct for any imbalance 
between the two detected signals \citep{tra00}. Such a dataset requires approximately five minutes of observing time.

\subsection{Visibility Measurement}

Methods of extraction of visibilities from these data are discussed in detail in 
Paper~II \citep{bru04} and generally follow the procedures described by 
Benson, Dyck, \& Howell (1995). For the data obtained here, the correlation $\nu$, measured 
from fringe amplitude, is treated as the proxy for visibility $V$. Algorithms were developed 
for seeking the interference fringe in each scan by locating the maximum excursion from 
mean intensity within the scan. Identification of false features in each scan can be a 
problem at low SNRs, and several methods were applied to guard against this. These include 
the tracking of the location of the fringe center, which should wander through successive 
scans in a continuous fashion, and calculating a SNR for each detected fringe based upon 
the amplitude of the peak in the power spectrum for that fringe. When a detected fringe is 
considered to be real, the amplitude of that fringe is determined by mathematical fitting 
of the fringe, and the amplitude of the fit is adopted as the correlation $\nu$. This means 
that the entire fringe packet, not just the maximum in the packet, contributes to the 
measurement of $\nu$. The sequence of fringe scans in a given dataset is subdivided by time 
into 15 subsets from which 15 mean values of $\nu$ are calculated, and the final value of 
$\nu$ and its standard error associated with a given dataset are taken as the mean and 
standard deviation arising from the 15 subsets.

The conversion of a correlation measured from fringe amplitude to a visibility 
is achieved by dividing the correlation of the calibrator star obtained by time 
interpolation between calibrator observations immediately before and after the target star. 
This is the standard practice in optical interferometry to compensate for the time varying
transfer function arising primarily from atmospheric fluctuations. It assumes a known value 
for the angular diameter of the calibrator whose predicted visibility can then be multiplied 
by the ratio of the target to calibrator correlations to obtain the visibility of the target star.

Estimating the visibility of the calibrator star (in this case HD 83362) is a critical factor 
in any interferometric diameter analysis. We have obtained a reliable estimate of the diameter of
the calibrator by utilizing the relationship 
existing among the parameters angular diameter, effective temperature, and surface flux. 
Thus, we have calculated an estimated angular diameter for the calibrator star HD 83362 
by fitting to a spectral energy distribution selected according to spectral type from a 
template compiled by Pickles (1998) to the available broadband photometry, 
particularly in the near-infrared \citep{gez93, cut03}. The resulting angular 
diameter for HD 83362 is 0.516$\pm$0.032 mas (UD), and we adopted this value 
for the calibration of visibilities for Regulus and the check star HD 88547.
The error in the estimated diameter for the calibrator star is $\pm6.2\%$, 
and this propagates into the measurements for the check star to produce $\pm0.9\%$ 
and $\pm2.5\%$ errors in calibrated visibility for baselines of 200~m and 300~m, respectively. 
For the larger diameter of Regulus, these errors are reduced to $\pm0.5\%$ and $\pm2.2\%$. 
The errors in derived diameters resulting from the uncertainty in the diameter of the 
calibrator are $\pm0.8\%$ for the check star and $\pm0.5\%$ for Regulus.  The overall errors 
in these determinations are significantly larger and dominated by the random error in 
visibility measurements. The uncertainty in the adopted diameter for the calibrator 
has a second order, and therefore negligible effect, on interferometric determination of 
relative shape and relative surface brightness variation.

This approach has led to 69 measurements of calibrated visibilities for Regulus and 40 such 
visibilities for HD 88547. Those values are summarized in Tables 1 and 2 and are available
on the CHARA website in the standard optical interferometry FITS format (Pauls et al. 2004). The distributions in 
the projected baseline plane are shown in Figure~1. These data are shown plotted in Figure~2 along 
with the best fits of uniform disk diameters using the standard equation for fitting $V(B)$ to obtain 
$\Theta_{UD}$ (cf. equation 3.41 of Traub 2000).  These fits yield $\Theta_{UD,HD88547}$~=~ 1.29$\pm$0.07 mas 
and $\Theta_{UD,Regulus}$~=~1.47$\pm$0.12 mas. It is apparent from the standard deviations 
of these values and simple inspection of Figure~2 that, while the diameter fit to the check 
star is reasonable and in good agreement with the diameter predicted from the available 
photometry, the fit to Regulus is much poorer with large systematic residuals. 
This immediately suggests that the interferometry of Regulus cannot be explained 
in terms of a simple uniform disk model.

\placefigure{fig1}  

\placefigure{fig2}  

\placetable{tab1}   
 
\placetable{tab2}   
 
\subsection{Modelling Visibilities Geometrically}

The interferometric analyses of the rapid rotators Altair \citep{bel01} and Achernar \citep{sou03} 
discussed the shapes of these stars on the basis of uniform disk diameter values 
associated with each $V^2$ prior to subjecting their data to more sophisticated models. 
Using this same approach as a starting point, we have selected observations made at the longest baselines 
that are clustered in several position angle regions as indicated in Figure~1. 
There are 6, 5, and 6 measurements of $V$ respectively in these clusters, and they yield 
UD diameters for Regulus of 1.413$\pm$0.024 mas, 1.514$\pm$0.028 mas, and 
1.328$\pm$0.028 mas for the three mean position angles $129\fdg1$, $181\fdg3$, 
and $251\fdg6$, respectively. A similar treatment of observations of the check star HD~88547 
results in UD diameters of 1.233$\pm$0.012 mas, 1.260$\pm$0.057 mas, and 1.246$\pm$0.015 mas 
at position angles $128\fdg8$, $206\fdg8$, and $253\fdg4$. These diameters are shown plotted in 
Figure~3 which clearly suggests that the check star is round while Regulus exhibits 
an elongated shape whose major axis is oriented along a position angle roughly in the 
range of 150$^\circ$ to 180$^\circ$.

\placefigure{fig3}  

We next fitted the visibility data with a series of elliptical models of constant surface brightness,
which we label as UE fits to distinguish from UD models.  These models 
have the advantage that their Fourier transform is just an Airy function, elongated according 
to the axial ratio \citep{bor99}.  These fits are a natural extension to the UD models 
done traditionally for (circular) stellar diameters.  Indeed, the fact that UE models are, 
like UD fits, Airy functions is why one may reasonably take the approach of the previous paragraph 
for a star whose shape is approximated by an ellipsoid. This three-parameter approach yields  
a best-fit UE model with $R_{\rm minor}/R_{\rm major} = 0.845\pm0.029$, 
a mean diameter of $1.42\pm0.04$~mas (or $R_{\rm minor} = 0.651$, $R_{\rm major} = 0.771$~mas), 
a position angle of the short axis of the ellipse of $84\fdg90 \pm 2\fdg4$ 
(measured eastward from north), and a reduced chi-square value of 3.41. 
These results indicate that the star is clearly non-spherical in shape, 
and in the following section we develop a physical model for the star so 
that we can make a realistic comparison between the predicted rotationally 
distorted shape and the observed visibilities.

\section{Spectroscopic Constraints on the Physical Parameters of Regulus}       

Rapid rotation in stars like Regulus has two immediate consequences for 
a physical description of the star.  First, the star will become oblate, and thus 
we need a geometrical relationship for the star's radius as a function of angle from the pole.  
Secondly, the surface temperature distribution will also become a function of co-latitude, 
usually assumed to be cooler at the equator than at the poles.  Reliable estimates of the 
predicted stellar flux, spectral features, and angular appearance
in the sky will then be functions of nine parameters:
mass $M$, polar radius $R_p$, polar effective temperature $T_p$, 
equatorial velocity $V_e$,
inclination angle of the pole to the line of sight $i$, 
position angle $\alpha$ of the polar axis from the north celestial pole 
through the east at the epoch of observation,
gravity darkening exponent $\beta$ that defines the amount of equatorial
cooling \citep{col91}, distance to the star $d$, and interstellar extinction. 
In principle one can make a grid search through these parameters 
to find the best-fit of the interferometric visibilities for 
different baselines projected onto the sky \citep{bel01}, but 
in the case of Regulus there are a number of spectral observations 
available that can help to reduce significantly the probable range in 
these parameters.  Here, we discuss how the spectral flux distribution
and the appearance of several key line profiles can be analyzed to 
form these additional constraints that we will apply to the analysis 
of the interferometric visibility data. 

We developed a physical model for the star by creating a 
photospheric surface grid of 40000 area elements. 
We assume that the shape of the star is given by the 
Roche model (i.e., point source mass plus rotation) 
for given equipotential surfaces based on the 
ratio of equatorial to critical rotational velocity 
$V_e/V_c$ \citep{col91}.  The effective 
temperature is set at each co-latitude according to the 
von Zeipel law \citep{zei24}, in which the temperature varies 
with the local gravity as $T(\theta)=T_p (g(\theta)/g_p)^\beta$
and $\beta$ is normally set at 0.25 for stars with radiative
envelopes \citep{cla03}.  The model integrates the specific intensity
from all the visible surface elements according to their projected area, 
temperature, gravity, cosine of the angle of the surface normal to 
the line of sight ($\mu$), and radial velocity.  The intensity spectra were 
calculated over the full range of interest in steps of $2000^\circ$ in 
temperature, $0.2$~dex in $\log g$, and 0.05 in $\mu$.  
The spectra were computed using the code {\it Synspec} 
\citep{lan03} and are based on solar abundance atmospheres models by 
R.\ L.\ Kurucz\footnote{\url{http://kurucz.harvard.edu/}}
(for a chosen microturbulent velocity of 1~km~s$^{-1}$). 
The model incorporates both limb darkening and gravity darkening
and provides reliable line profiles even in cases close to 
critical rotation \citep*{tow04}.  Each model spectrum is 
convolved with an appropriate instrumental broadening function 
before comparison with observed spectra.  The code is also used to 
produce monochromatic images of the star in the plane of the sky. 

We begin by considering the full spectral flux distribution to 
develop constraints on the stellar radius and temperature. 
We plot in Figure~4 observations 
of the flux of Regulus binned into wavelength increments of 
the same size as adopted in the Kurucz model flux spectrum shown. 
The measurements for the extreme ultraviolet are from \citet{mor01}, 
and the far- and near-ultraviolet spectra are from the archive
of the {\it International Ultraviolet Explorer (IUE) Satellite}
(spectra SWP33624 and LWP10929, respectively).  The optical 
spectrophotometry is an average of the observations from 
\citet{ale96} and \citet{bor03}. 
The near-infrared fluxes are from the IR magnitudes given by 
\citet{bou91} that we transformed to fluxes using the 
calibration of \citet{coh92}.  
We initially fit this distribution using simple flux models 
for a non-rotating star, and the fit illustrated in Figure~4
uses the parameters  
$T_{\rm eff}=12250$~K, $\log g = 3.5$, $E(B-V)=0.005$, 
and a limb-darkened angular diameter of $1.36\pm 0.06$~mas. 
These parameters agree well with those from \citet{cod76}
based upon a full flux integration and the angular diameter 
from intensity interferometry.  

\placefigure{fig4}  

We used the fit of the spectral energy distribution in $K$-band
as the basis of an observational constraint on the integrated 
flux for our rotating star model.  The comparison was made by 
calculating the $K$-band flux for a given model and then 
adjusting the flux according to the distance from the 
{\it Hipparcos} parallax measurement of Regulus, $\pi=42.09\pm0.79$ mas or 
$d=23.8\pm0.4$ pc \citep{per97}, and the
negligible extinction derived from $E(B-V)$ \citep{fit99}.  
We require that modeled and observed fluxes agree to better than $1\%$ 
(the error associated with the fit of the absolute observed flux in the
$K$-band).  This constraint primarily affects the selection 
of the polar temperature (which sets the flux) and polar radius 
(which together with the adopted distance sets the angular 
area of the source). 

The next constraint is provided by the shape of the 
hydrogen H$\gamma$ $\lambda 4340$ line profile as 
shown in Figure~5.  This observed spectrum is 
from the spectral library of \citet{val04}, and it has 
resolving power of $\lambda/ \triangle \lambda = 4900$. 
This Balmer line grows in equivalent 
width with declining temperature through the B-star range, 
and it is wider in dwarfs than in supergiants because of 
pressure broadening (linear Stark effect).   
Thus, fits of the profile are set by the adopted model temperature 
and surface gravity distribution over the visible hemisphere of the star.  

\placefigure{fig5}  

Our final constraint is established by the rotational line 
broadening observed in the 
\ion{Mg}{2} $\lambda 4481$ line as shown in Figure~6.
This spectrum is the sum of 30 spectra obtained in 1989 April 
with the Kitt Peak National Observatory 0.9-m Coud\'{e} Feed Telescope, 
and it has a resolving power of $\lambda/ \triangle \lambda = 12400$. 
This spectral region also includes the weaker metallic lines of 
\ion{Ti}{2} $\lambda 4468$, \ion{Fe}{2} $\lambda 4473$, and
\ion{Fe}{1} $\lambda 4476$, which become prominent in cooler 
A-type spectra.  All these lines have intrinsically narrow intensity 
profiles, and the broadening we observe is due primarily to 
Doppler shifts caused by stellar rotation.  A fit of the \ion{Mg}{2} $\lambda$4481 
profile leads directly to the projected equatorial rotational velocity, 
$V\sin i$, which for a given inclination then defines the actual 
rotational speed and stellar deformation.  Note that we exclude 
the nearby \ion{He}{1} $\lambda 4471$ line from this analysis, 
since it is possible that Regulus belongs to the class of He-weak stars 
that account for about one quarter of the B-stars with 
temperatures similar to that of Regulus \citep{nor71}. 

\placefigure{fig6}  

We assume throughout our analysis that the optical and $K$-band flux
originates solely in the stellar photosphere.  This assumption needs 
verification because many rapidly rotating B-type stars develop 
large equatorially confined disks (spectroscopically identified as
Be stars; \citealt{por03}) and such disks can contribute a 
large fraction of the $K$-band flux \citep{ste01}.
The appearance of excess IR flux from a disk is always 
accompanied by the development of an emission feature in the 
Balmer H$\alpha$ profile \citep{ste01}, so we have searched 
for any evidence of H$\alpha$ emission in spectra obtained 
contemporaneously with the interferometric observations. 
We show in Figure~7 the average H$\alpha$ profile in the 
spectrum of Regulus formed from 11 spectra obtained with the 
KPNO Coud\'{e} Feed Telescope on 2004 October 13 -- 16
(less than six months after our most recently collected interferometric data). 
This spectrum has a resolution of $R = \lambda/ \triangle \lambda =
10600$ and a signal-to-noise ratio of 500 per pixel in the continuum. 
It appears completely free of any excess emission such as that
shown in Figure~7 for two Be stars observed at the same time as 
Regulus (HD~22780 and HD~210129, both of comparable spectral 
classification, B7~Vne).   We also show in Figure~7 the
photospheric H$\alpha$ profile of a slowly rotating B-star, HD~179761, 
which we obtained from the spectral atlas of \citet{val04}. 
This star has a temperature $T_{\rm eff} = 13000$~K and
a gravity $\log g = 3.5$ \citep{smi93} that are comparable to 
the surface mean values for Regulus.  However, it has a very small
projected rotational velocity, $V\sin i = 15$ km~s$^{-1}$
\citep*{abt02}, so we artificially broadened the profile by convolution with 
a rotational broadening function with $V\sin i = 317$ km~s$^{-1}$
(see below) and a linear limb darkening coefficient of 0.30
\citep{wad85} in order to match approximately the line broadening in the spectrum 
of Regulus.  The good agreement between the spun-up version of 
the H$\alpha$ profile of HD~179761 and that of Regulus indicates 
that no excess disk emission is present.  
We have also examined spectra ($R = 14000$; S/N = 200 -- 500)
made with the CHARA 1.0-m equivalent aperture 
Multi-Telescope Telescope (MTT) \citep*{bar02} on six nights 
(1999 January 19, March 17, 2000 February 29, April 16, 20, 29), and 
none of these showed any evidence of emission.  Furthermore, another
KPNO Coud\'{e} Feed spectrum from 2000 December and 21 HEROS spectra \citep*{ste00} made
between 2001 May and 2002 May (kindly sent to us by Dr.\ S.\ \v{S}tefl) are free of
H$\alpha$ emission. These contemporary observations of no emission are generally consistent
with the record in the literature.  For example, \citet{sle78} used 
Regulus as a standard emission-free and non-varying star in a survey for
H$\alpha$ variations among Be stars during the period 1975 December to
1977 June. There is only one report of marginal H$\alpha$ emission in the 
spectrum of Regulus in 1981 February \citep{sin82}, but the observations
at hand indicate that there was no disk at the time of the CHARA Array
observations, so that the $K$-band visibilities can be analyzed reliably in terms of a 
photospheric model alone.

\placefigure{fig7}  

We constructed a series of stellar models that consistently meet 
all of these observational constraints for a grid of assumed 
values of stellar inclination $i$ and gravity darkening coefficient $\beta$. 
The method was an iterative approach in which we first established 
the projected rotational broadening from fits of the \ion{Mg}{2}
$\lambda 4481$ line (dependent only on $\beta$ over the inclination 
range of interest) and then progressively approximated the 
polar temperature and radius to find models that met both the $K$-band flux and 
H$\gamma$ profile constraints.  Note that once the temperature 
and radius are set, the H$\gamma$ fit provides us with the gravity ($\log g$)
and hence the final parameter, stellar mass.  The results of these 
solutions for the $i$ and $\beta$ grid are listed in Table~3, which includes the fitted projected
rotational velocity (column 3), the ratio of equatorial to critical velocity (column 4),
the polar and equatorial radii (columns 5 and 6), the mass (column 7), and the polar and
equatorial temperatures (columns 8 and 9). 

\placetable{tab3}   
 
The first derived quantity is the projected rotational velocity 
$V\sin i$, which depends mainly on the adopted value of gravity 
exponent $\beta$.  With larger $\beta$, the equatorial regions 
are darker and contribute less in the extreme line wings. 
Consequently, the best fits accommodate a slightly larger value 
of $V\sin i$ at the nominally accepted value of $\beta=0.25$
compared to the case of no gravity darkening, $\beta=0$. 
Fitting errors admit a range of $\pm3$ km~s$^{-1}$ in the 
derived $V\sin i$ value.  We find, however, that the fits 
of \ion{Mg}{2} $\lambda 4481$ become much worse for models 
close to critical rotation, which confirms the inclination 
range of $60^\circ < i < 90^\circ$ found by \citet{sto87} 
based on similar profile studies.  The estimates of stellar 
radii and mass are almost independent of the choice of 
$i$ and $\beta$, and it is only the range in stellar temperature
between the equator and pole that increases with $\beta$. 
Note, however, that the geometric mean of the equatorial and 
polar temperatures is almost the same in all the models, 
which results from matching the integrated $K$-band flux 
(reflecting the average temperature across the visible hemisphere). 
The internal errors of the fitting procedure (exclusive of any 
systematic errors associated with the model atmospheres and fluxes) 
lead to errors of $\pm1\%$ in the radii and $\pm7\%$ in mass.

\section{Fits of Interferometric Visibility Using Physical Models}

All of the models in Table~3 make acceptable fits of the 
spectroscopic data, and to further discriminate between them 
we turn to the model predictions about the interferometric 
visibilities.  Here we discuss how to predict the visibility 
patterns associated with the physical models and how the 
observed visibilities can be used to evaluate the remaining 
parameters, in particular the inclination and the gravity darkening
exponent.  We first explore the goodness of fit for the 
full grid of models shown in Table~3 and we determine what parameter
set minimizes the reduced chi-square for the whole set and subsets of
the visibilities.  We then compare the results with an independent 
approach based upon a guided search of the multi-parameter space. 

We created $K$-band surface intensity images of the stellar
disk for each case in Table~3 and made a two-dimensional Fourier transform 
of the image for comparison with the interferometric visibility \citep{dom02}.   
The modulus of the Fourier transform relative to that at frequency zero
is then directly comparable to the observed visibility at a 
particular coordinate in the $(u,v)$ spatial frequency plane 
(defined by the ratio of the projected baseline to the effective 
wavelength of the filter).   

The predicted variation in visibility as a function of baseline 
is shown in Figure~8 for three selected models and 
for two orientations: the top set of three plotted lines corresponds to 
an interferometric baseline parallel to the polar (minor) axis 
while the lower set corresponds to the equatorial (major) axis. 
The CHARA Array observations correspond to baselines between 
188 and 328~m, so we are sampling the main lobe of visibility. 
Model A corresponds to the case with $i=90^\circ$ and $\beta=0.25$
(see Table~3).  The visibility drops much more rapidly in 
the direction along the major axis compared to the minor axis, 
and consequently the variations in interferometric visibility 
in different directions in the $(u,v)$ plane set the orientation 
of the star in the sky.  Model B is also based on the nominal 
gravity darkening exponent value of $\beta=0.25$, but the 
inclination is $i=70^\circ$ in this case.   There are subtle 
differences in the pole to equator axis visibilities from model A 
that result from the greater rotational distortion and smaller polar 
radius in model B.   Finally, model C shows the $i=90^\circ$ 
case but with no gravity darkening ($\beta=0$).  Now the equatorial 
flux is easily seen out to the extreme approaching and receding limbs, 
and the star appears larger in this dimension (resulting in lower visibility). 
These models show that although the differences due to stellar 
orientation in the sky are seen at all baselines, the smaller 
differences related to stellar inclination and gravity darkening 
are most apparent at longer baselines.  Indeed, future observations 
with even longer baselines (or shorter wavelengths) that reach into the 
second lobe would strongly discriminate between these models. 

\placefigure{fig8}  

We made a two-parameter fit to match the predicted and observed 
visibilities of Regulus.  The first parameter is the position 
angle of the minor (rotation) axis in the sky $\alpha$. 
The second parameter is an angular 
scaling factor, the ratio of best fit distance to the adopted 
distance from the {\it Hipparcos} parallax.  In principle, 
the combination of the $K$-band flux and adopted temperature model
leads to a predicted angular size (or stellar radius for a given 
distance), but the observed errors in absolute flux and distance 
will lead to a predicted angular size that may be marginally 
different from the actual angular size.  The use of the scaling parameter 
provides a simple way to both check the prediction and adjust the spatial frequency 
scaling to best fit the observed visibilities.  The results of these 
fits are given in Table~3 which lists the minor axis angular radius (column 10),
major axis angular radius (column 11), the position angle $\alpha$ (column 12), 
the scaling factor $d/d(Hipparcos)$ (column 13), and
the reduced chi-square of the fit $\chi^2_\nu$ (column 14;
for $N=69$ measurements and $\nu=2$ fitting parameters).  
We estimate that the errors in position angle are $\pm2\fdg8$ and in scaling
factor are $\pm1\%$ (based upon the increase in $\chi^2_\nu$ with
changes in these parameters). We see that the scaling factors are generally less than
$2\%$ different from unity, well within the error budget from
the comparable errors in the estimates of the $K$-band flux,
parallax, and effective wavelength of the CHARA $K^{\prime}$ filter.

The best overall fit is found for the rotational model with 
$i=90^\circ$ and $\beta=0.25$, and we show the predicted image of the stellar disk 
in the sky and its associated visibility 
in the $(u,v)$ plane for this model in Figure~9.  The upper part of the right panel 
shows the visibility as a gray-scale intensity and the polar axis 
as a dotted line (along the longer dimension in the visibility 
distribution).  Black squares mark the positions in the $(u,v)$ plane 
corresponding to the CHARA interferometric observations.  The lower
panel shows the point symmetric $(u,v)$ plane, and there each observation 
is assigned a gray-scale intensity based upon the normalized residual,
$(V_{\rm obs}-V_{\rm cal})/\sigma_V$.  Note that the poor fits are the 
most apparent in this representation, but that there is no evidence 
of any systematic trends in the residuals with position in the $(u,v)$ plane.
The relative close proximity in the $(u,v)$ plane of measurements that
are above and below the predicted visibility suggests that the 
outstanding differences are due to measurement error rather than 
deficiencies in the models.  The fact that the minimum chi-square 
is $\chi^2_\nu=3.35$ rather than the expected value of unity indicates 
that our internal estimates of error in visibility (from the scatter 
within a sequence of fringe scans) may not adequately represent the total error budget.
However, it is well known that the temporal power spectrum of atmospheric fluctuations
has a steep slope. Since the calibration cycle has a much lower frequency than the scan to scan
cycle, it is natural that the noise estimate from the latter underestimates the former.

\placefigure{fig9}  

We showed above in Figure~8 that the interferometric visibilities 
at all baselines provide a strong constraint on the position angle $\alpha$, 
and we illustrate this striking dependence in Figure~10.  
This series of panels shows the normalized residuals from the model 
predictions for the $i=90^\circ$ and $\beta=0.25$ case as a function 
of baseline.  The sky orientation of the $K$-band image of the star is shown to the right 
of each panel.  The residuals for all baselines clearly 
obtain a well defined minimum for $\alpha=85\fdg5$ ({\it third panel from top}). 
Figure~11 shows the reduced chi-square as a function of $\alpha$ for 
both the entire set of measurements ({\it solid line}) and the long 
baseline measurements only ($B_{\rm proj} > 270$ m; {\it dotted line}).
Both sets yield a consistent minimum, and the long baseline data 
are particularly sensitive to the position angle orientation. 

\placefigure{fig10}  

\placefigure{fig11}  

The visibility constraints on the inclination and gravity darkening 
exponent are less pronounced but still of great interest. 
We show in Figure~12 the reduced chi-square as a function of 
the gravity darkening exponent $\beta$ for a series of $i=90^\circ$ 
model fits.  The best fit occurs at $\beta=0.25$ for the full set 
of observations, and this is also the value derived from gravity darkening 
studies of B-stars in eclipsing binary stars \citep{cla03}.   
The formal $1\sigma$ error limit yields an acceptable range 
from $\beta = 0.12$ to 0.34, but $\beta=0$ (no gravity darkening)
can only be included if we extend the range to the $99\%$ confidence level.
However, recall from Figure~8 that most of the sensitivity to 
gravity darkening is only found at longer baselines and 
especially at those along the polar axis.  Thus, we also show in 
Figure~12 the value of reduced chi-square for the $i=90^\circ$ 
solutions in two subsets:
measurements with baselines greater than 270~m (31 points) 
and those with a position angle close to the polar axis 
orientation (6 points from MJD~53103.2 -- 53104.2; see Table~1). 
Both sets (especially the latter, polar one) show much larger
excursions and indicate $\beta$ values close to 0.25. 
Note that we did not try models with $\beta > 0.35$ since such 
models have equatorial temperatures that are cooler than the lower 
limit of our flux grid.  The most sensitive polar subset helps establish
more practical error limits, and we adopt $\beta = 0.25 \pm 0.11$ as 
our best overall estimate.  This analysis demonstrates that the inclusion 
of gravity darkening is key to fitting the CHARA data 
in the most sensitive part of the visibility curve. 

\placefigure{fig12}  

The greater sensitivity of the long baseline and polar data is also apparent 
in the fits with differing inclination angle.  We show in Figure~13 
the run of reduced chi-square with inclination for models with 
$\beta=0.25$.  A shallow minimum at $i=90^\circ$ is found using all 
the data, but the same minimum is much more convincingly indicated 
in fits of the long baseline and polar data alone.   
The $1\sigma$ error limit of the latter sample admits a range of $78^\circ - 90^\circ$. 
\citet{hut77} also derived an inclination of $i=90^\circ$ for Regulus through 
a comparison of ultraviolet and optical photospheric line widths.  

\placefigure{fig13}  

It is interesting to compare the rotation axis vector direction, 
determined here by combination of the angles $i$ and $\alpha$, with the 
space velocity vector direction. Hipparcos proper motion and parallax 
results \citep{per97} combined with a new radial velocity value of 
$V=+7.4\pm2.0$~km~s$^{-1}$, derived from the analysis described above 
of the \ion{Mg}{2} $\lambda 4481$ line, provide the necessary input for making 
this comparison. The angular values representing the space velocity that 
correspond to our measurements of $i=90^{\circ}\pm15^{\circ}$ and 
$\alpha=85\fdg5\pm2\fdg8$ are $i_s=75\fdg2\pm4\fdg0$ and $\alpha_s=271\fdg1\pm0\fdg1$. 
Because $\alpha$ is inherently ambiguous by $180^{\circ}$, 
$\alpha-\alpha_s=-5\fdg6\pm2\fdg8$ and $i-i_s=14\fdg8\pm15\fdg5$. 
This suggests that Regulus is moving very nearly pole-on through space.

We have checked these models through a comparison with an independent 
code based upon the scheme first outlined by \citet{bel01}, using a new generation of that 
code that conducts its comparisons in Fourier space rather than in image space.  This method uses 
a different numerical realization of the geometry of the star, but one that 
is based on the same Roche approximation adopted above.  There are six 
key parameters involved in this method (polar radius, $i$, 
$\alpha$, $\beta$, rotation speed, and, albeit with low sensitivity, polar temperature) 
that define the projection of the stellar surface on the sky. 
This sky projection was then run 
through a two-dimensional Fourier transform, and a reduced chi-square was calculated from a comparison of the
observed and model values for $V^2$.
A multi-dimensional optimization code was then utilized to derive the best 
solution, a process that took typically 500 iterations \citep{pre92}.
The results are compared with those from fitting a uniform ellipsoid (\S2.2)
and from the spectroscopically constrained fits (Table~3) in Table~4. 
We see that all three methods agree on the orientation of the polar 
axis in the sky, and both the grid search and spectroscopically constrained 
approaches find the same results for the angular sizes and inclination. 
The only discrepancy concerns the value of $\beta$, 
and the lower value derived from the grid search scheme is probably 
the result of the insensitivity to this parameter of the majority of 
the measurements in the whole sample.   We place more reliance on 
the results of the sensitive polar sample in the spectroscopically 
constrained fit (Fig.~12).

\placetable{tab4}   

In summary, we find that infrared interferometric measurements of Regulus 
over a wide range of position angle are consistent with parameters derived 
from spectroscopic criteria, and determine additional parameters which are not 
available from spectroscopy, particularly including the position angle of the 
rotation axis on the sky $\alpha$, the inclination of that axis to the 
line of sight $i$, and the gravity darkening coefficient $\beta$. 
Our adopted results and their errors \placetable{tab3} are summarized in Table~5. 
Our physical models for this rapidly rotating star indicate that Regulus has an 
equatorial radius that is $32\%$ larger than the polar radius. 
Its rotation period is 15.9 hours, which corresponds to an equatorial 
rotation speed that is $86\%$ of the critical break-up velocity.   
Fits of the CHARA observations require the presence of significant 
gravity darkening with an equatorial temperature that is only $67\%$ 
of the polar temperature.  

\placetable{tab5}   

\section{Conclusions}   

We have shown in a series of increasingly complex models, ultimately tied to 
physical parameters strongly constrained by spectroscopy, that Regulus exhibits 
features expected for a rapidly rotating star of its spectral type, namely oblateness 
and gravity darkening. Geometric fits to fringe visibility, first with discrete 
position angle determinations of uniform disk diameters and then
with an ellipsoidal model, clearly show the marked rotational oblateness of the star. 
When we couple the visibilities to models that incorporate parameters to which high 
resolution spectroscopy is sensitive, we find mutual consistency between the 
interferometric and spectroscopic results. Indeed, we believe that the combination 
of these complementary astrophysical probes - interferometry and spectroscopy - 
provides the best means for exploiting new high spatial resolution measurements 
from such instruments as the CHARA Array.

Our observations provide the first interferometric evidence of gravity darkening 
in rapidly rotating stars. Furthermore, the CHARA Array results offer the first 
claim of gravity darkening in a star that is not a known member of an eclipsing 
binary system \citep{cla03}. The agreement between the measurement of the angular 
diameter from the CHARA visibilities and that based upon the $K$-band flux provides 
an independent verification of the infrared flux method for estimating angular 
diameters first introduced by \citet{bla77}, and it indicates that the B-star 
fluxes predicted by line blanketed, LTE atmospheres models by R.\ L.\ Kurucz agree 
with the observed angular diameter and infrared flux within the observed errors.

New interferometric observations of this kind offer the means to 
determine how close the most rapidly rotating stars are to their 
critical rotation speeds, and this may help solve the longstanding 
problem of the nature of mass loss in the rapidly rotating Be stars, 
for example \citep{por03}. These observations may also finally allow us to 
test models of interior structure and evolution for massive rotating stars \citep{end79,heg00,mey00}.   
We show in Figure~14 the position of Regulus in the theoretical 
Hertzsprung-Russell diagram based upon the surface integration of $\sigma T^4$ 
to estimate the luminosity and upon an average temperature derived from the 
luminosity and surface area of our best-fit model.  We also show the zero-age 
to terminal-age main sequence evolutionary paths for non-rotating 3 and $4 M_\odot$ 
stars based upon the work of \citet{sch92}.   The shaded region represents the 
main sequence evolutionary parameters for a star with the mass and errors in mass 
we derive from the joint spectroscopic-interferometric analysis. The position of 
Regulus in this diagram shows that the star 
is overluminous for its mass as predicted for rotating stars that 
evolve to higher luminosity as fresh H is mixed into their 
convective cores \citep{heg00,mey00}.  If Regulus is placed on an isochrone 
without consideration of this excess luminosity, the age of the star is overestimated. 
This is the likely explanation for the apparent discordance in age that has been noted 
for the $\alpha$~Leo A and B components \citep{ger01}. The new era of long baseline 
interferometry offers us the means to probe the evolution of rotating stars as has 
never before been possible. 

\placefigure{fig14}  

\section{Acknowledgements}

This research has been supported by National Science Foundation grants AST--0205297 and AST--0307562. Additional support has been received from the Research Program Enhancement program administered by the Vice President for Research at Georgia State University. Portions of this work were performed at the California Institute of Technology under contract with the National Aeronautics and Space Administration. We thank CHARA Array Operator P. J. Goldfinger for her care in obtaining many of these observations.

As this is the first scientific paper from the CHARA Array, a project whose origin dates back to the formal establishment of CHARA at Georgia State University in 1984, it is appropriate to acknowledge the many individuals and organizations responsible for bringing the dream of this facility into reality. The College of Arts and Sciences has generously supported CHARA from its conceptual beginnings in 1983 when Dean Clyde Faulkner agreed to establish a research center with the goal of building a facility for high angular resolution astronomy. The present Dean, Lauren Adamson, graciously continues to support the center in increasingly difficult financial times. Georgia State President Carl Patton has been an enthusiastic supporter of CHARA, having visited the Array site on numerous occasions. Tom Lewis, Vice President for External Affairs, has often accompanied President Patton to California and has been a strong proponent for the project. We are particularly indebted to Cleon Arrington, former Vice President for Research and Sponsored Programs, who worked with tireless enthusiasm with the CHARA director to raise matching funds to build the Array. The exceptional nurturing and substantial backing given by the administration of Georgia State University is deeply appreciated.

We also wish to acknowledge the continuing support provided by administrative service elements of the University during the years of construction of the Array. We particularly thank Albertha Barrett, now Assistant Vice President for Research, and her staff in research administration for their patience and expertise in dealing with grants and contracts related matters. This project has involved thousands of procurements, and we acknowledge David Bennett, Larry McCalop, and Howard Hopwood of the University's purchasing department for their exceptional efforts to ensure that CHARA got what it needed when it needed it. The expertise of Charles Hopper, manager of the physics and astronomy shop, has been critical to the materialization of the hundreds of custom designed and fabricated parts that comprise the many subsystems of the Array. CHARA Site Manager Bob Cadman has ably served as our first line of defense on any number of matters, from groundbreaking to the present time. Finally, the devotion of Alexandra ``Sandy'' Land, CHARA's Business Manager, has been a key ingredient to seeing this effort finished on schedule and within budget.

The National Science Foundation has provided substantial funding for preliminary and detailed design and ultimately for construction of the CHARA Array through NSF grant AST--9414449. The kind support and careful oversight provided by Wayne Van Citters, Kurt Weiler, Benjamin Snavely, and James Breckinridge, successive directors of the Advanced Technology and Instrumentation Program, is gratefully acknowledged. 

The W. M. Keck Foundation provided funding to expand the Array from a five- to a six-telescope instrument and to enhance our beam combining capability. We thank Maria Pellegrini and Mercedes Talley of the Keck Foundation for making this support possible.

We thank Kenneth Ford of the David and Lucile Packard Foundation for his role in providing funds that capped the University's matching obligation to the NSF.

Jack Kelly, Georgia State physics alumnus, kindly donated funds in support of an exhibit hall attached to CHARA's main operations building on Mount Wilson.

Several individuals worked closely with CHARA during the planning and design years, and we acknowledge the valuable contributions made by Allen Garrison and William Robinson of the Georgia Tech Research Institute and also by William Hartkopf, our colleague for many years at Georgia State who now continues CHARA's original tradition of binary star speckle interferometry at the U.S. Naval Observatory.

We thank Robert Jastrow, former director of the Mount Wilson Institute, for inviting us to explore Mount Wilson as a possible site for the CHARA Array and for assisting in many ways to ease the complications of obtaining site access. Once on Mount Wilson, we were fortunate to have the services of Eric Simison, president of Sea West Enterprises, who became not only our prime contractor but also a core member of our design and engineering team. We also acknowledge Terry Ellis, former District Ranger for the Angeles River Ranger District of the Angeles National Forest for his kind guidance in our efforts to fulfill NEPA and Department of Agriculture guidelines in locating on Mount Wilson.

Ingemar Furenlid served on the astronomy faculty at Georgia State from 1982 until he passed away on 1994 February 11. He was a respected colleague and dear friend to several of us and at all times a hearty cheerleader for CHARA. We deeply regret that he is not among the co-authors of this publication, and we dedicate this first CHARA Array paper to his memory.


\setcounter{figure}{0}
\clearpage

\begin{figure}
\begin{center}
\includegraphics[angle=0,height=12cm]{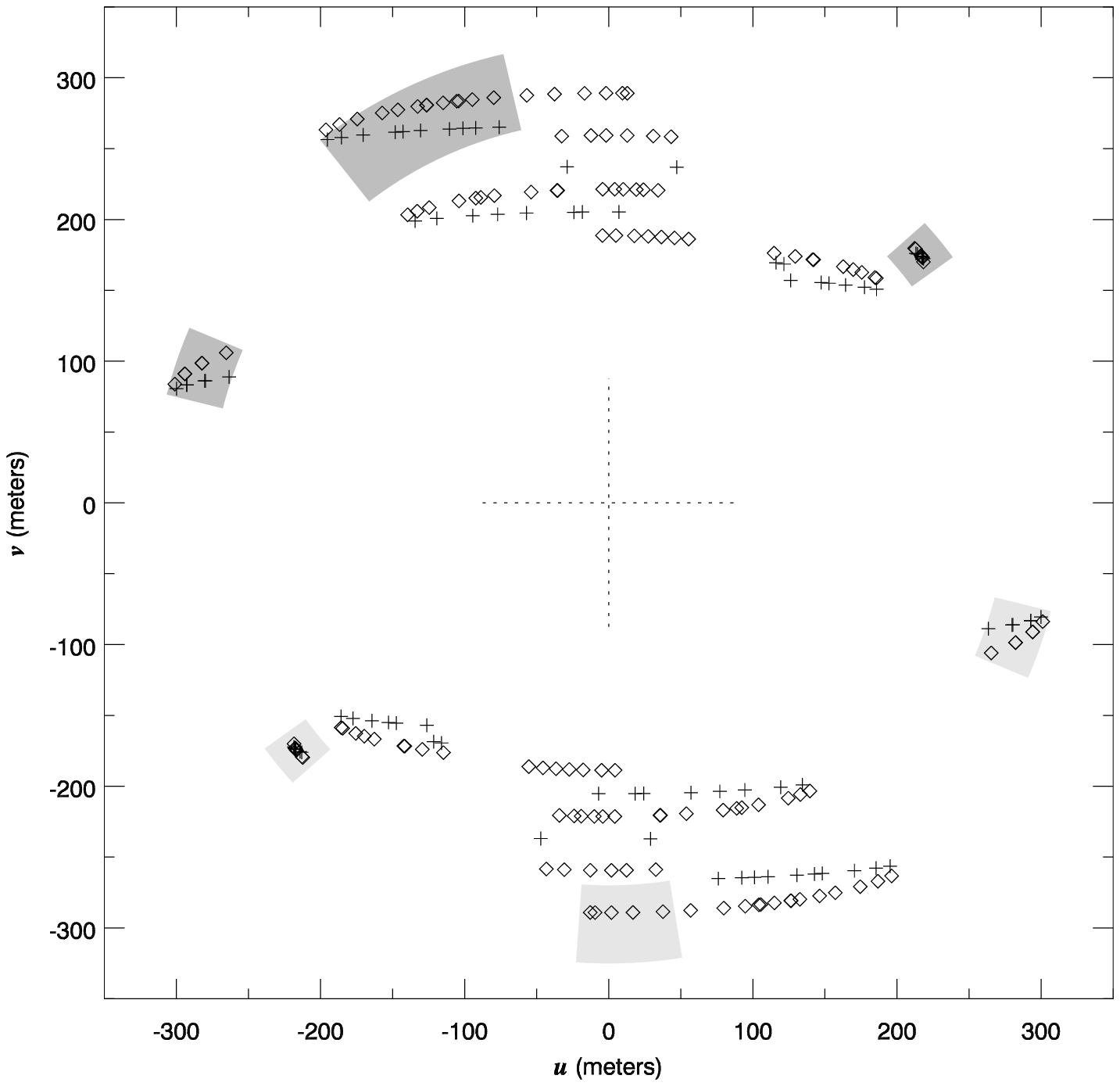}
\end{center}
\caption{The baseline coverages for Regulus ({\it open diamonds}) and
the check star HD~88547 ({\it plus signs}) are shown in the $(u,v)$ plane above.
The lighter shaded regions in the lower half indicate those measurements
of Regulus used for a position angle dependent estimate of diameter
(see Fig.~3), while the darker shaded regions in the upper half show
the same for measurements of the check star.}
\label{fig1}
\end{figure}

\begin{figure}
\begin{center}
\plotone{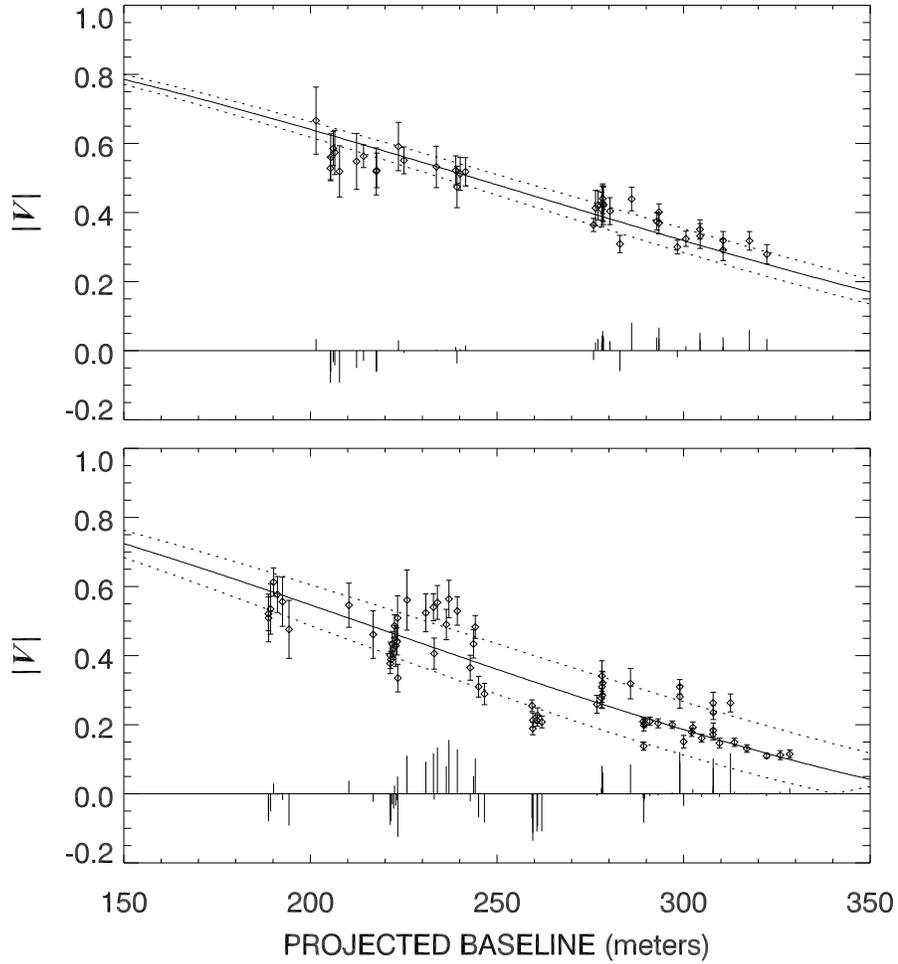}
\end{center}
\caption{The best-fit uniform disk diameter fits to the calibrated
visibilities are shown for the check star HD~88547 ({\it top panel}) and Regulus
({\it bottom panel}). The data for Regulus are clearly poorly fit by a uniform disk model.}
\label{fig2}
\end{figure}

\begin{figure}
\begin{center}
\includegraphics[angle=90,height=12cm]{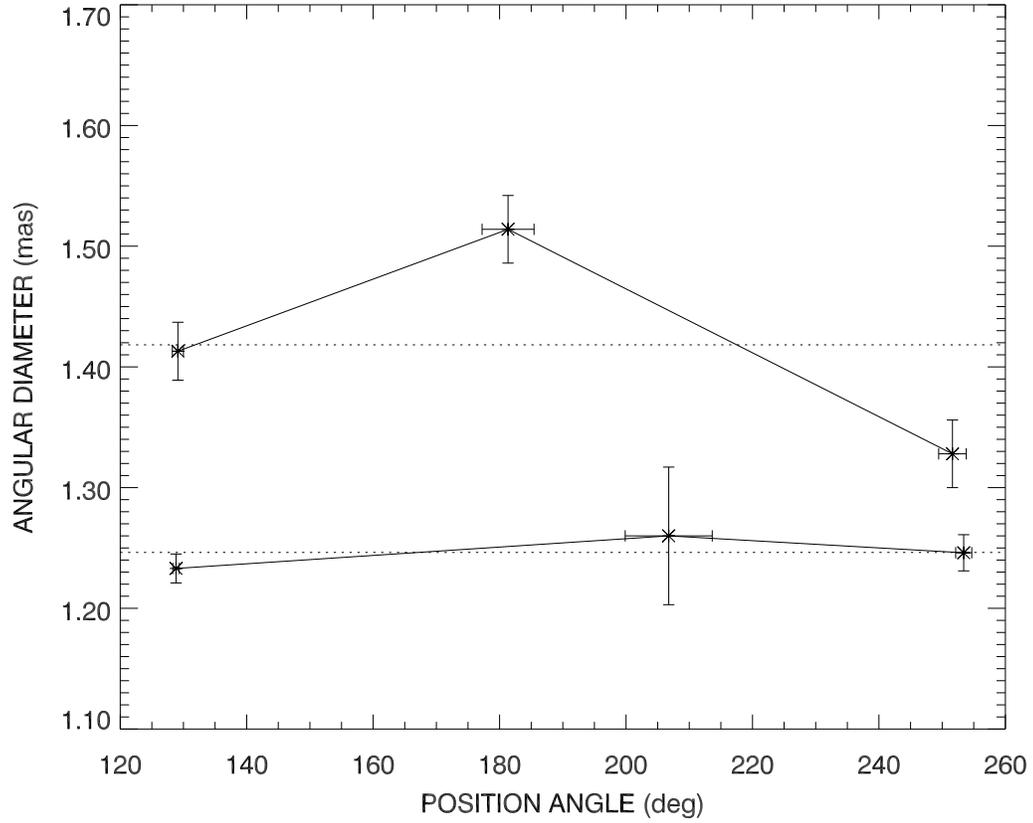}
\end{center}
\caption{Uniform disk diameters calculated from the longest baseline
data in the vicinity 
of three position angles indicated in Fig.~1 are shown for the check
star HD~88547 ({\it lower values}) and Regulus ({\it upper values}). 
The dotted lines show the respective mean values of these 3 sets in each
case. These results imply that the check star is round while Regulus is not.}
\label{fig3}
\end{figure}

\begin{figure}
\begin{center}
\includegraphics[angle=90,height=12cm]{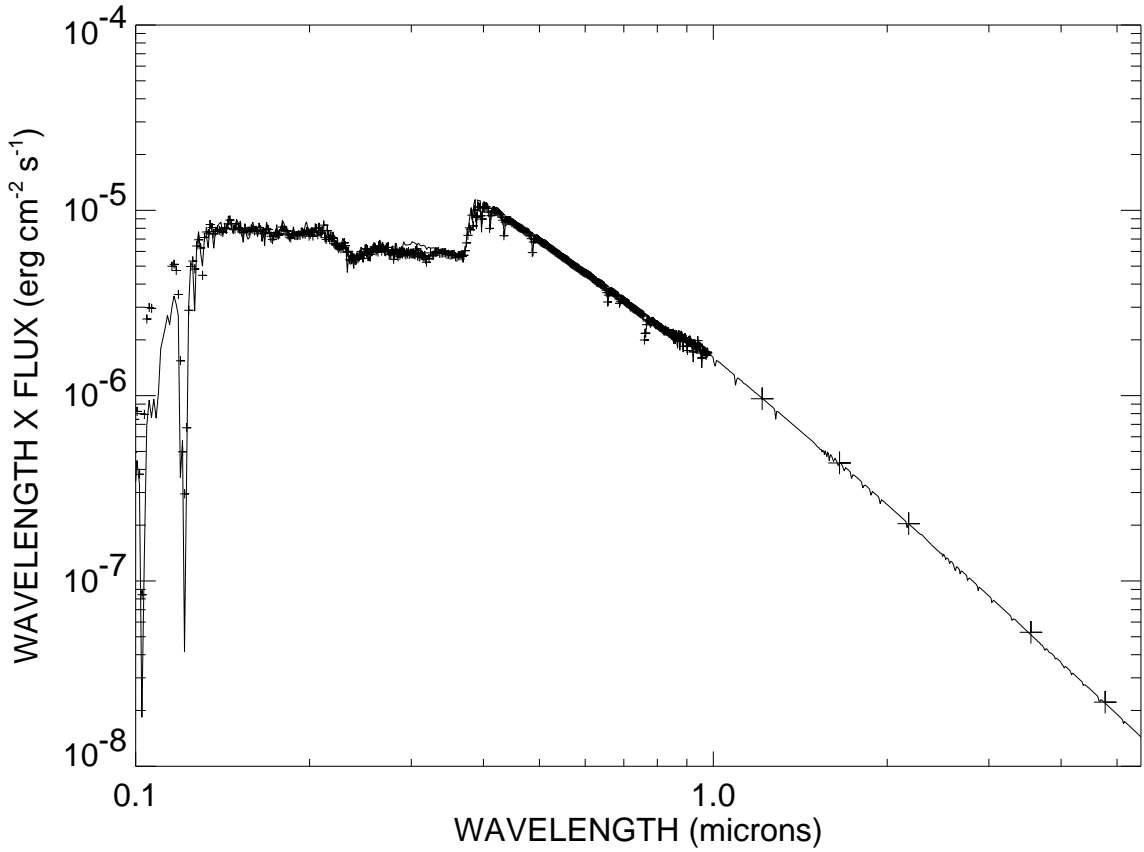}
\end{center}
\caption{The full spectral flux distribution of Regulus. 
Smaller plus signs indicate the short wavelength observations which are
binned to the same resolution as the model distribution ({\it solid line})
while the larger plus signs indicate fluxes from infrared magnitudes.}
\label{fig4}
\end{figure}

\begin{figure}
\begin{center}
\includegraphics[angle=90,height=12cm]{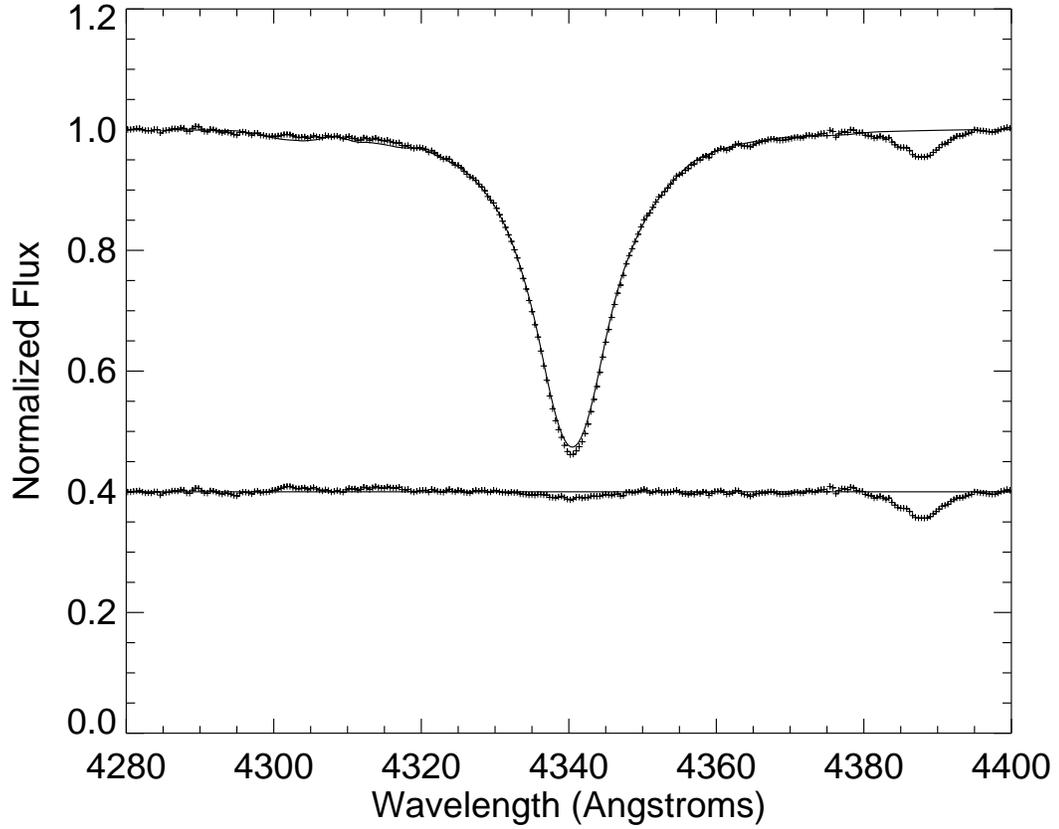}
\end{center}
\caption{The observed ({\it plus signs}) and model ({\it solid line}) 
profiles for H$\gamma$ in the spectrum of Regulus.  The plot
corresponds to a model with a gravity darkening exponent $\beta=0.25$ and
an inclination $i=90^\circ$.  The lower plot shows the residuals from 
the fit (note that the \ion{He}{1} $\lambda 4387$ line that was not 
included in the calculation).}
\label{fig5}
\end{figure}

\begin{figure}
\begin{center}
\includegraphics[angle=90,height=12cm]{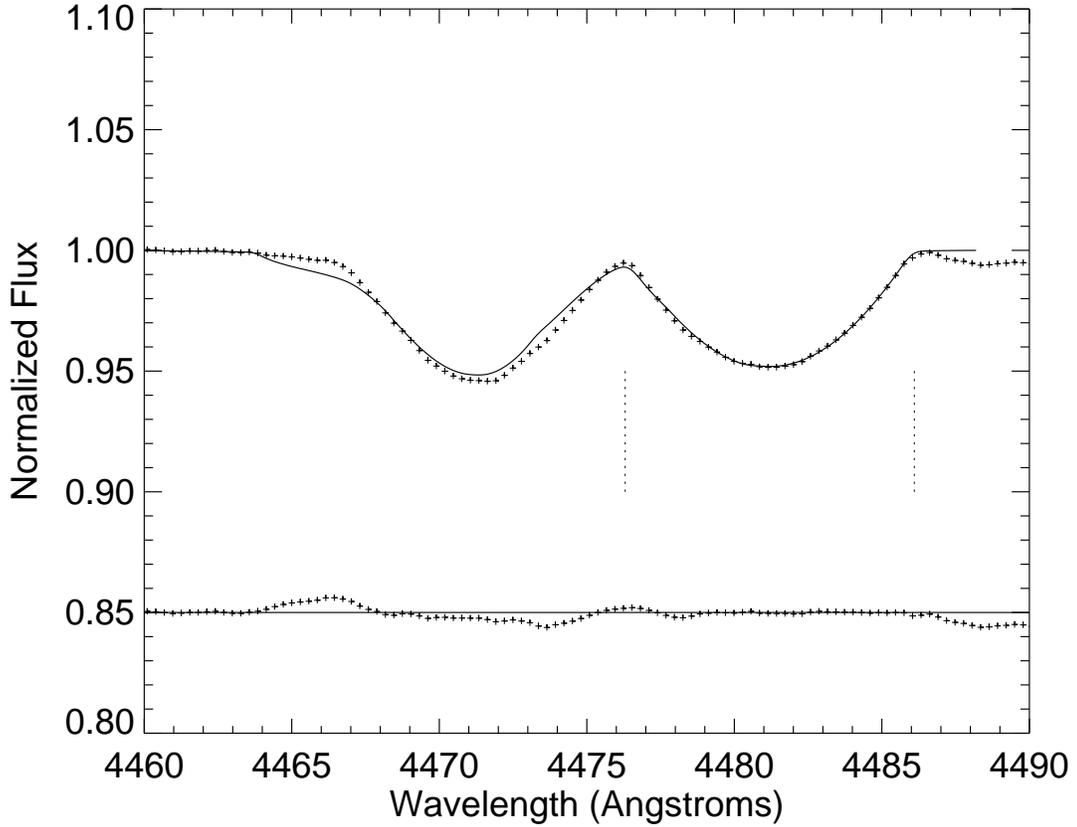}
\end{center}
\caption{The observed ({\it plus signs}) and model ({\it solid line}) 
profiles for \ion{Mg}{2} $\lambda 4481$ in the spectrum of Regulus.  
The plot corresponds to a model with $i=90^\circ$ and $\beta=0.25$.
Residuals from the fit are shown below. 
Note that the \ion{He}{1} $\lambda 4471$ line and other weaker lines 
were not included in the fit of projected rotational velocity (which was 
based on the interval between the two vertical dashed lines.)}
\label{fig6}
\end{figure}

\begin{figure}
\begin{center}
\includegraphics[angle=90,height=12cm]{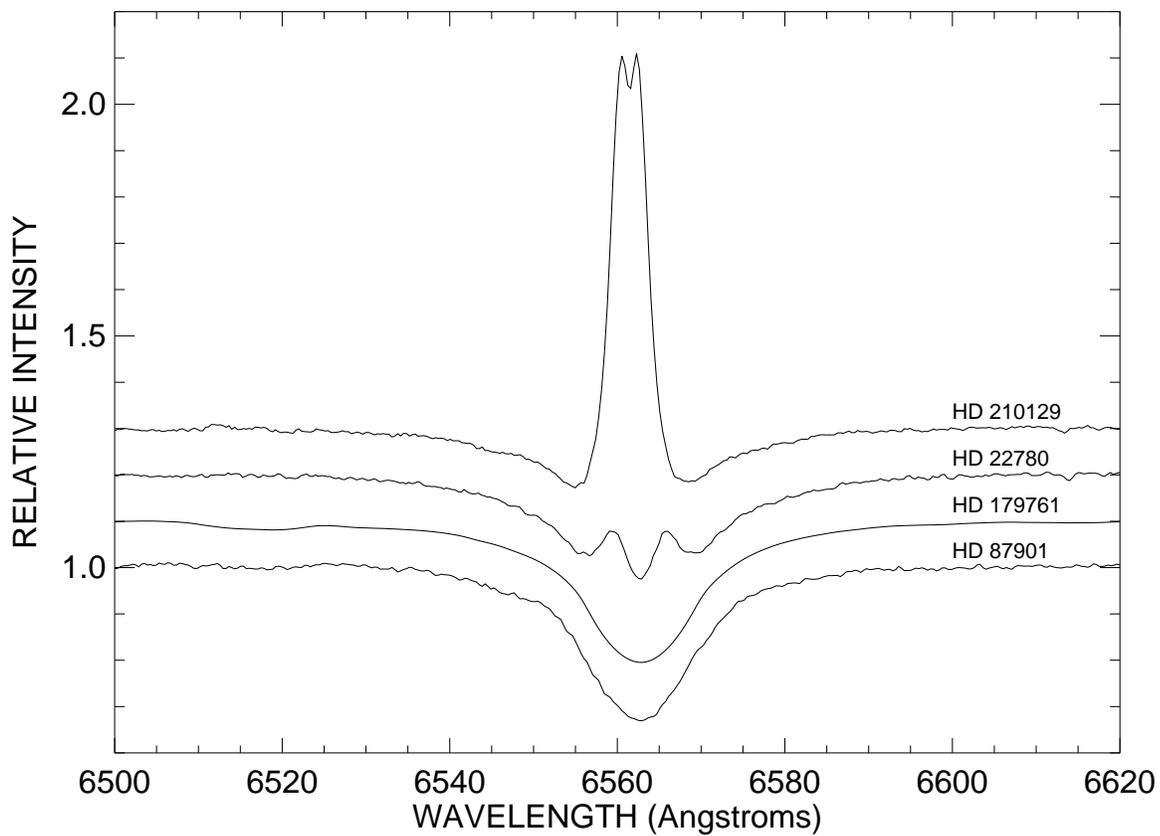}
\end{center}
\caption{A plot of the H$\alpha$ profile in the spectrum of Regulus 
({\it bottom}; HD~87901).  This is an average of eleven spectra made with 
the KPNO Coud\'{e} Feed Telescope from 2004 October 13 -- 16.  
There is no evidence of the kind of disk H$\alpha$ emission that is 
observed in rapidly rotating Be stars such as 
HD~22780 ({\it second from top}; B7~Vne) 
and HD~210129 ({\it top}; B7~Vne).  Also shown is the photospheric 
H$\alpha$ profile of the slowly rotating star HD~179761
({\it third from top}; B8~II-III), which we artificially broadened 
to match the rotational broadening of Regulus.   The close 
agreement between the photospheric line of HD~179761 and
that of Regulus indicates that no disk emission is present.}
\label{fig7}
\end{figure}

\begin{figure}
\begin{center}
\includegraphics[angle=90,height=12cm]{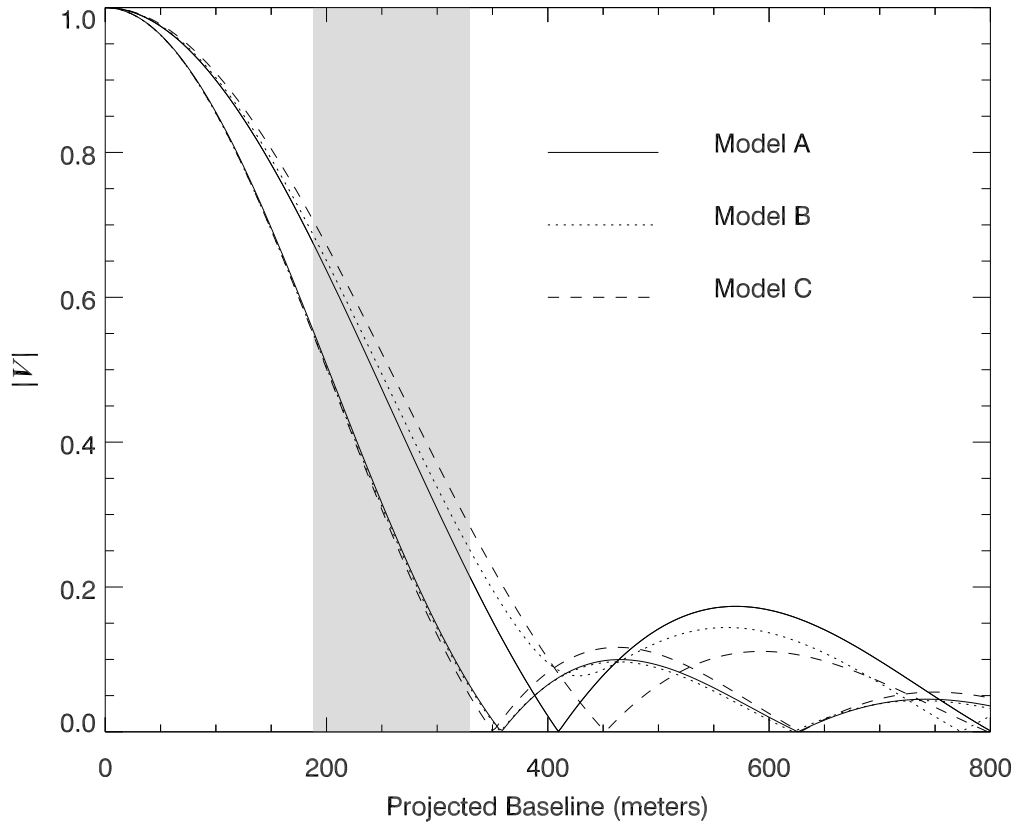}
\end{center}
\caption{Predicted visibility variations with baseline for three
rotation models and two sky orientations.  The upper group corresponds to a
baseline parallel to the minor (rotational) axis in the sky while the lower group 
corresponds to a baseline parallel to the major axis.   The visibility
curves are shown for the cases of 
$i=90^\circ, \beta=0.25$ ({\bf A}; {\it solid lines}),
$i=70^\circ, \beta=0.25$ ({\bf B}; {\it dotted lines}), and 
$i=90^\circ, \beta=0$ ({\bf C}; {\it dashed lines}). 
The spatial frequency for a given baseline is shown for a
filter effective wavelength of $2.1501$ $\mu$m.
The shaded region indicates the baseline range of the 
CHARA Array observations.}
\label{fig8}
\end{figure}

\begin{figure}
\plotfiddle{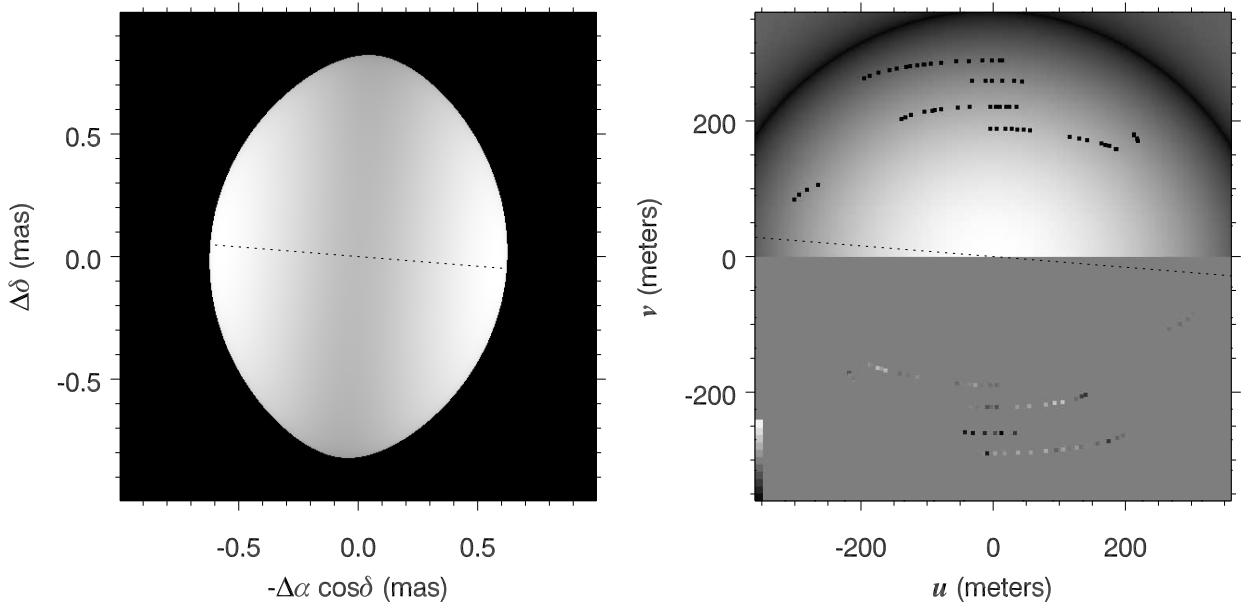}{0.0cm}{90}{329}{600}{-50}{-2000}
\caption{A $K$-band image of the star in the sky ({\it left}) and its
associated Fourier transform visibility pattern in the $(u,v)$ plane ({\it right}). 
In both cases north is at the top and east is to the left. 
The dotted, black line indicates the direction of the rotational 
axis for this $i=90^\circ$, $\beta =0.25$, and $\alpha= 85\fdg5$ model.
The upper panel of the visibility figure ({\it right}) shows a 
grayscale representation of the visibility and the positions of the 
CHARA measurements ({\it black squares}).  The lower panel shows
the normalized residuals from the fit as a gray scale intensity square 
against a gray background in a point symmetric representation of 
the $(u,v)$ plane.  The legend at lower left shows the intensities corresponding to 
normalized residuals from $-5$ ({\it black}) to $+5$ ({\it white}). 
Note that the best fit points appear gray and merge with the background.}
\label{fig9}
\end{figure}

\begin{figure}
\epsscale{1.0}
\plotone{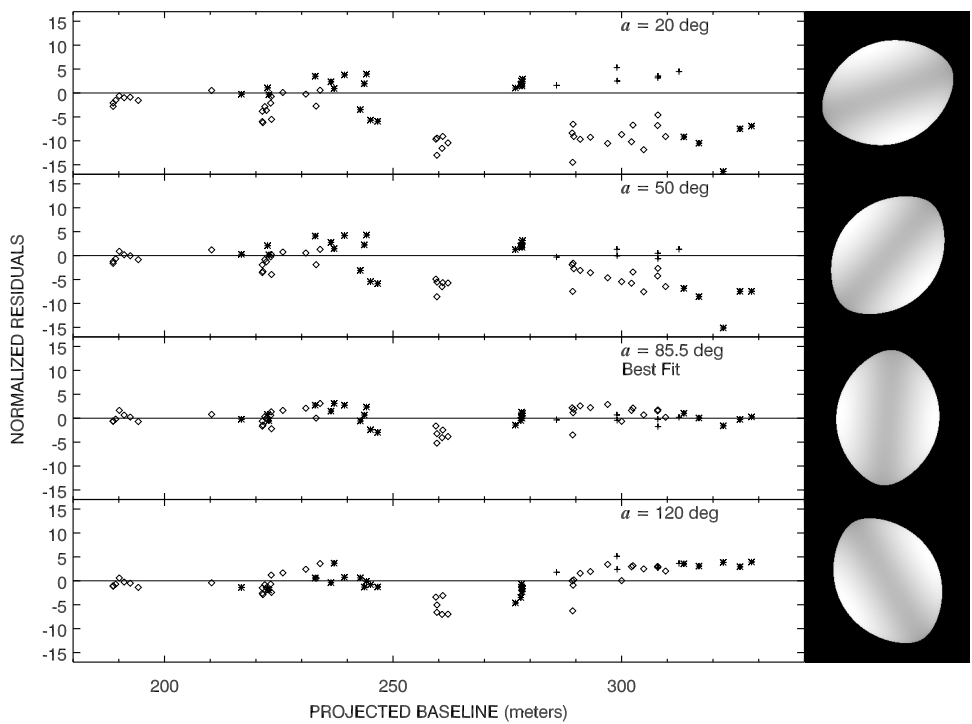}
\caption{Normalized visibility residuals as a function of baseline. 
Each panel shows the residuals for the model star with $i=90^\circ$, 
$\beta=0.25$, and a position angle $\alpha$ as indicated (and
illustrated at right). The residuals are clearly minimized at the best fit value of
$\alpha=85\fdg5$ ({\it third panel from top}). Plus signs indicate measurements in the 
$(u,v)$ plane within $30^\circ$ of the rotation axis (6 points), 
diamonds indicate those within $30^\circ$ of the equator (40 points), 
and asterisks indicate the others at intermediate angles (23 points).}
\label{fig10}
\end{figure}

\begin{figure}
\begin{center}
\includegraphics[angle=90,height=12cm]{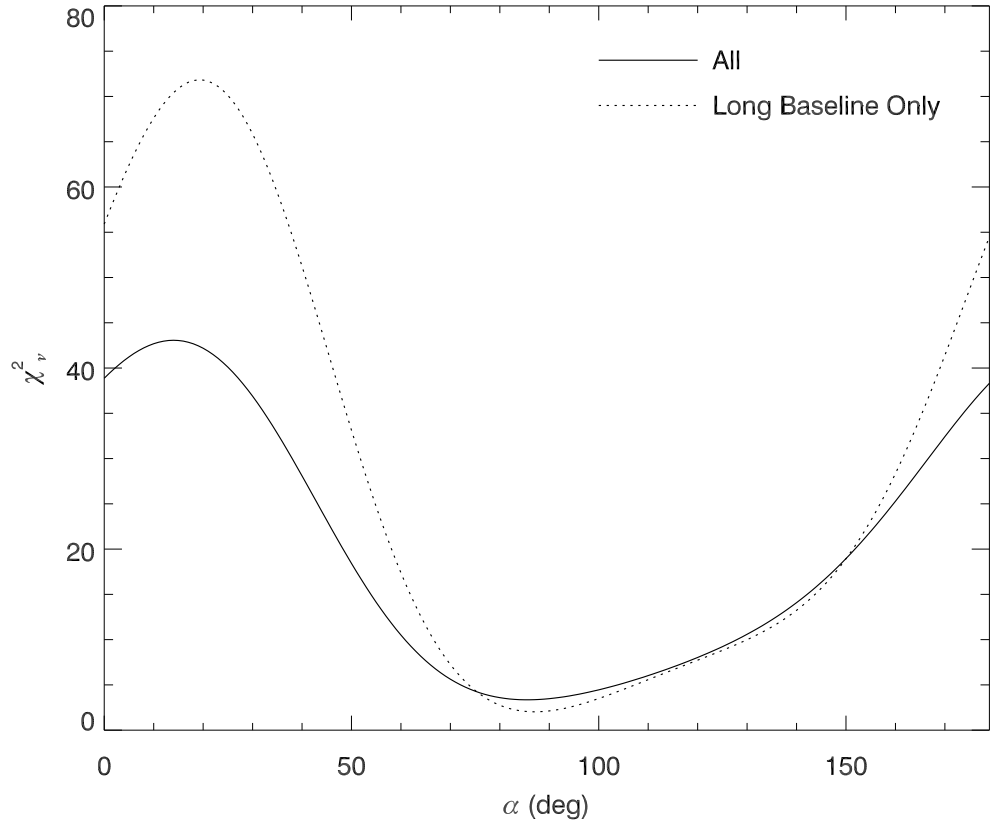}
\end{center}
\caption{A plot of the reduced chi-square $\chi^2_\nu$ of the visibility fits 
as a function of position angle $\alpha$ (for $i=90^\circ$ and $\beta=0.25$). 
The solid line shows the reduced chi-square for whole sample while the 
dotted line shows the same for the long baseline data only.}
\label{fig11}
\end{figure}

\begin{figure}
\begin{center}
\includegraphics[angle=90,height=12cm]{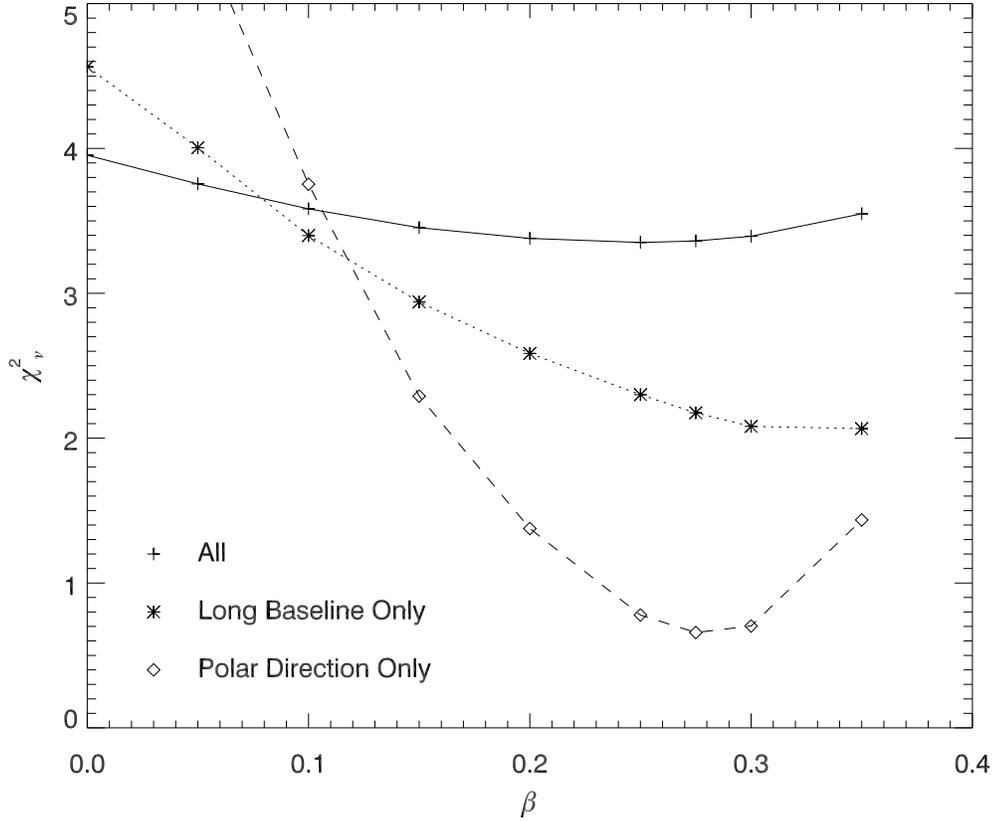}
\end{center}
\caption{A plot of the reduced chi-square $\chi^2_\nu$ of the visibility
fits as a function of gravity darkening exponent $\beta$ (for $i=90^\circ$). 
The solid line shows the reduced chi-square for whole sample (69
points). The dotted line shows the reduced chi-square of the same fits for the 
long baseline data only (31 points) while the dashed line shows the same
for those long baseline data with a position angle near the orientation
of the polar axis (the 6 points most sensitive to the selection of
$\beta$).  All these samples indicate a gravity darkening exponent near the 
predicted value of $\beta=0.25$.}
\label{fig12}
\end{figure}

\begin{figure}
\begin{center}
\includegraphics[angle=90,height=12cm]{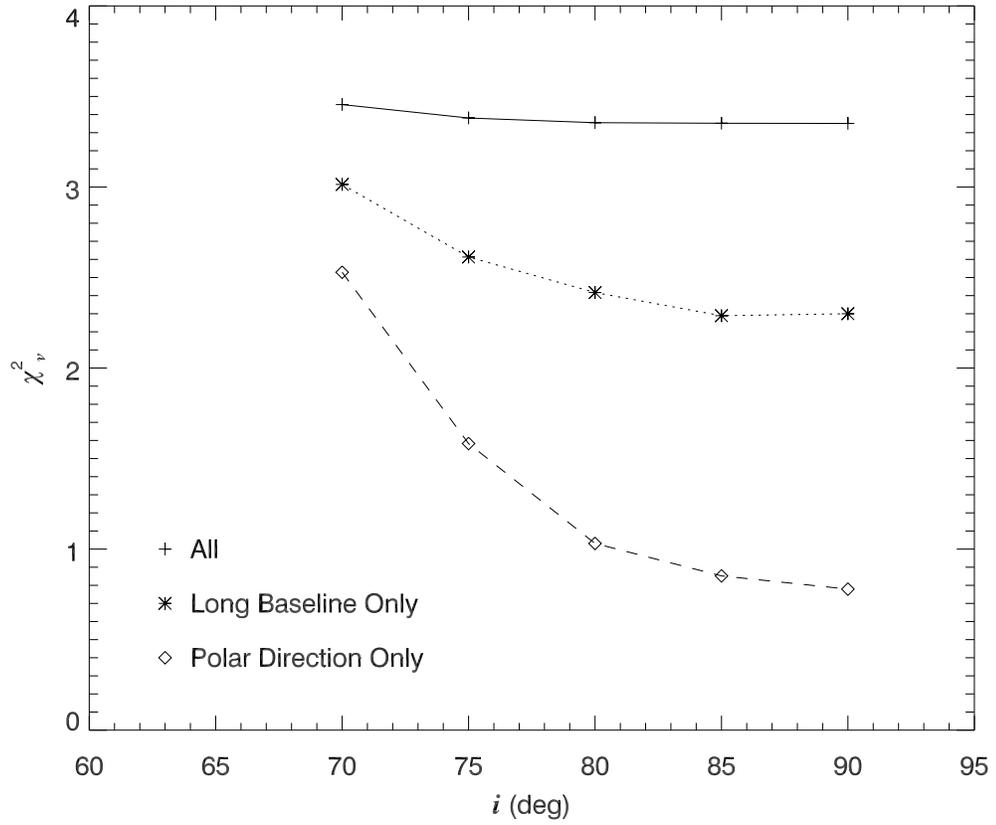}
\end{center}
\caption{A plot of the reduced chi-square $\chi^2_\nu$ of the visibility
fits as a function of the rotation axis inclination angle $i$ (for
$\beta=0.25$). The different lines correspond to the same samples as shown in Fig.~12.}
\label{fig13}
\end{figure}

\begin{figure}
\begin{center}
\includegraphics[angle=90,height=12cm]{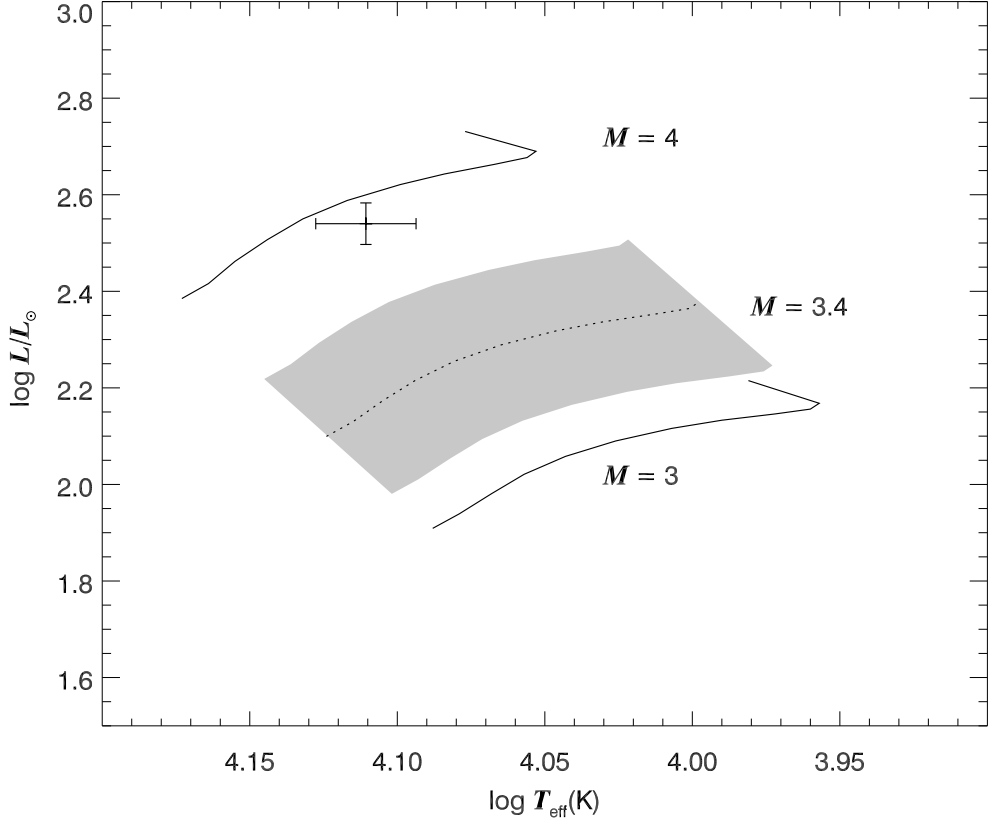}
\end{center}
\caption{The Hertzsprung-Russell diagram for Regulus.  
The two solid lines show the track from zero age to 
terminal age main sequence ({\it left to right}) for non-rotating 
stars with initial masses of 3 and $4M_\odot$ \citep{sch92}, 
while the dotted line and surrounding shaded area show the 
predicted region for a non-rotating star with the derived mass 
of Regulus and its associated error.  The single point at 
$L/L_\odot = 347$ and $<T_{\rm eff}> = 12901$~K (from a 
surface integration of our numerical model) is well above the 
predicted position, indicating that the star is overluminous 
for its mass in comparison to models for non-rotating stars.} 
\label{fig14}
\end{figure}


\clearpage
\begin{deluxetable}{cccc}
\tablecolumns{4}
\tabletypesize{\scriptsize}
\tablewidth{0pc}
\tablecaption{Interferometric Measurements of Regulus (HD 87901)}
\tablehead{
\colhead{} & \colhead{Projected} & \colhead{Baseline} & \colhead{} \\
\colhead{} & \colhead{Baseline} & \colhead{Position Angle} & \colhead{Calibrated}\\
\colhead{MJD} & \colhead{(m)} & \colhead{(deg)} & \colhead{Visibility}}
\startdata
53074.188	&	188.76	&	181.33	&	0.520	$\pm$	0.049	\\
53074.200	&	188.77	&	178.49	&	0.508	$\pm$	0.069	\\
53074.215	&	189.31	&	174.62	&	0.535	$\pm$	0.073	\\
53074.227	&	190.11	&	171.71	&	0.613	$\pm$	0.041	\\
53074.238	&	191.17	&	168.98	&	0.576	$\pm$	0.052	\\
53074.250	&	192.51	&	166.29	&	0.556	$\pm$	0.071	\\
53074.263	&	194.26	&	163.41	&	0.476	$\pm$	0.084	\\
53078.237	&	222.54	&	140.50	&	0.486	$\pm$	0.032	\\
53078.262	&	232.97	&	135.70	&	0.539	$\pm$	0.039	\\
53078.280	&	239.36	&	132.82	&	0.528	$\pm$	0.041	\\
53078.297	&	244.18	&	130.51	&	0.483	$\pm$	0.033	\\
53079.207	&	210.30	&	146.95	&	0.546	$\pm$	0.063	\\
53079.222	&	216.82	&	143.36	&	0.461	$\pm$	0.069	\\
53079.235	&	222.84	&	140.36	&	0.427	$\pm$	0.059	\\
53079.269	&	236.42	&	134.15	&	0.490	$\pm$	0.044	\\
53080.250	&	233.17	&	202.38	&	0.405	$\pm$	0.045	\\
53080.309	&	223.41	&	189.25	&	0.334	$\pm$	0.040	\\
53080.351	&	221.40	&	178.92	&	0.400	$\pm$	0.038	\\
53080.366	&	221.95	&	175.06	&	0.433	$\pm$	0.036	\\
53080.382	&	223.23	&	171.17	&	0.440	$\pm$	0.039	\\
53081.167	&	246.64	&	214.46	&	0.289	$\pm$	0.031	\\
53081.181	&	245.06	&	212.86	&	0.310	$\pm$	0.030	\\
53081.196	&	242.83	&	210.87	&	0.365	$\pm$	0.036	\\
53081.228	&	237.11	&	206.00	&	0.563	$\pm$	0.054	\\
53081.243	&	234.04	&	203.22	&	0.553	$\pm$	0.048	\\
53081.259	&	230.92	&	200.10	&	0.524	$\pm$	0.054	\\
53081.288	&	225.89	&	193.79	&	0.560	$\pm$	0.087	\\
53081.307	&	223.35	&	189.13	&	0.509	$\pm$	0.064	\\
53081.339	&	221.40	&	181.09	&	0.376	$\pm$	0.029	\\
53081.354	&	221.53	&	177.40	&	0.387	$\pm$	0.025	\\
53081.368	&	222.29	&	173.79	&	0.420	$\pm$	0.031	\\
53088.224	&	313.63	&	207.81	&	0.148	$\pm$	0.012	\\
53088.243	&	307.88	&	204.22	&	0.172	$\pm$	0.016	\\
53088.262	&	302.17	&	200.16	&	0.178	$\pm$	0.012	\\
53088.281	&	296.95	&	195.59	&	0.200	$\pm$	0.011	\\
53088.299	&	293.18	&	191.18	&	0.203	$\pm$	0.013	\\
53088.314	&	290.95	&	187.43	&	0.209	$\pm$	0.012	\\
53088.329	&	289.54	&	183.33	&	0.205	$\pm$	0.013	\\
53088.340	&	289.19	&	180.37	&	0.208	$\pm$	0.013	\\
53088.351	&	289.39	&	177.45	&	0.198	$\pm$	0.017	\\
53092.231	&	307.94	&	204.26	&	0.183	$\pm$	0.021	\\
53092.250	&	302.49	&	200.41	&	0.191	$\pm$	0.016	\\
53093.148	&	328.43	&	216.70	&	0.115	$\pm$	0.012	\\
53093.163	&	325.90	&	214.97	&	0.112	$\pm$	0.013	\\
53093.179	&	322.24	&	212.79	&	0.110	$\pm$	0.007	\\
53093.199	&	316.93	&	209.74	&	0.131	$\pm$	0.010	\\
53093.223	&	309.63	&	205.36	&	0.146	$\pm$	0.014	\\
53093.239	&	304.84	&	202.15	&	0.160	$\pm$	0.011	\\
53093.256	&	300.01	&	198.40	&	0.151	$\pm$	0.018	\\
53093.335	&	289.30	&	178.12	&	0.137	$\pm$	0.012	\\
53095.249	&	243.70	&	130.75	&	0.433	$\pm$	0.041	\\
53103.200	&	312.56	&	254.45	&	0.263	$\pm$	0.026	\\
53103.219	&	307.95	&	252.80	&	0.234	$\pm$	0.020	\\
53103.239	&	298.97	&	250.74	&	0.310	$\pm$	0.021	\\
53103.259	&	285.76	&	248.24	&	0.318	$\pm$	0.044	\\
53104.216	&	307.92	&	252.79	&	0.262	$\pm$	0.031	\\
53104.236	&	299.02	&	250.75	&	0.281	$\pm$	0.033	\\
53105.261	&	278.11	&	130.25	&	0.340	$\pm$	0.044	\\
53105.279	&	278.34	&	128.86	&	0.320	$\pm$	0.031	\\
53106.282	&	277.96	&	128.48	&	0.276	$\pm$	0.028	\\
53107.290	&	276.77	&	127.91	&	0.259	$\pm$	0.026	\\
53108.255	&	278.23	&	130.09	&	0.284	$\pm$	0.038	\\
53108.274	&	278.14	&	128.63	&	0.310	$\pm$	0.045	\\
53111.265	&	260.88	&	187.19	&	0.227	$\pm$	0.022	\\
53111.281	&	259.59	&	182.72	&	0.189	$\pm$	0.018	\\
53111.291	&	259.37	&	180.40	&	0.255	$\pm$	0.017	\\
53111.300	&	259.61	&	177.17	&	0.213	$\pm$	0.021	\\
53111.314	&	260.72	&	173.20	&	0.212	$\pm$	0.016	\\
53111.324	&	262.02	&	170.49	&	0.207	$\pm$	0.017	\\
\enddata
\end{deluxetable}

\clearpage

\begin{deluxetable}{cccc}
\tablecolumns{4}
\tabletypesize{\scriptsize}
\tablewidth{0pc}
\tablecaption{Interferometric Measurements of the Check Star (HD 88547)}
\tablehead{
\colhead{} & \colhead{Projected} & \colhead{Baseline} & \colhead{}\\
\colhead{} & \colhead{Baseline} & \colhead{Orientation} & \colhead{Calibrated}\\
\colhead{MJD} & \colhead{(m)} & \colhead{(deg)} & \colhead{Visibility}}
\startdata
53074.369	&	205.43	&	145.59	&	0.560	$\pm$	0.068	\\
53074.384	&	207.80	&	144.20	&	0.519	$\pm$	0.074	\\
53078.246	&	214.24	&	136.51	&	0.563	$\pm$	0.033	\\
53078.267	&	225.06	&	133.10	&	0.551	$\pm$	0.039	\\
53078.286	&	233.75	&	130.61	&	0.532	$\pm$	0.060	\\
53079.222	&	201.50	&	141.20	&	0.666	$\pm$	0.097	\\
53079.250	&	217.65	&	135.41	&	0.519	$\pm$	0.068	\\
53080.267	&	217.75	&	200.73	&	0.522	$\pm$	0.050	\\
53080.331	&	206.12	&	185.08	&	0.585	$\pm$	0.049	\\
53081.181	&	240.14	&	214.04	&	0.512	$\pm$	0.048	\\
53081.208	&	233.51	&	210.73	&	0.455	$\pm$	0.045	\\
53081.243	&	223.58	&	204.98	&	0.591	$\pm$	0.070	\\
53081.287	&	212.36	&	195.60	&	0.548	$\pm$	0.081	\\
53081.322	&	206.53	&	186.73	&	0.574	$\pm$	0.064	\\
53081.354	&	205.36	&	178.02	&	0.528	$\pm$	0.033	\\
53088.230	&	298.34	&	208.59	&	0.300	$\pm$	0.020	\\
53088.267	&	282.95	&	200.93	&	0.309	$\pm$	0.025	\\
53088.287	&	275.86	&	195.99	&	0.364	$\pm$	0.019	\\
53093.153	&	322.30	&	217.29	&	0.279	$\pm$	0.028	\\
53093.168	&	317.64	&	215.72	&	0.318	$\pm$	0.027	\\
53093.187	&	310.57	&	213.29	&	0.319	$\pm$	0.026	\\
53093.211	&	300.64	&	209.54	&	0.324	$\pm$	0.022	\\
53093.228	&	293.42	&	206.43	&	0.401	$\pm$	0.024	\\
53093.246	&	286.07	&	202.73	&	0.439	$\pm$	0.034	\\
53093.261	&	280.27	&	199.24	&	0.404	$\pm$	0.039	\\

53095.254	&	239.28	&	129.06	&	0.474	$\pm$	0.060	\\
53103.207	&	310.56	&	254.96	&	0.292	$\pm$	0.031	\\
53103.224	&	304.43	&	254.14	&	0.332	$\pm$	0.036	\\
53103.245	&	292.83	&	252.89	&	0.374	$\pm$	0.025	\\
53103.264	&	278.05	&	251.37	&	0.409	$\pm$	0.050	\\
53104.222	&	304.38	&	254.14	&	0.351	$\pm$	0.028	\\
53104.241	&	293.38	&	252.94	&	0.369	$\pm$	0.031	\\
53105.267	&	276.40	&	129.57	&	0.413	$\pm$	0.051	\\
53105.285	&	278.32	&	128.73	&	0.440	$\pm$	0.043	\\
53106.287	&	278.48	&	128.55	&	0.424	$\pm$	0.050	\\
53107.296	&	278.19	&	128.27	&	0.427	$\pm$	0.050	\\
53108.262	&	277.03	&	129.35	&	0.421	$\pm$	0.042	\\
53108.283	&	278.49	&	128.52	&	0.421	$\pm$	0.056	\\
53111.271	&	238.95	&	186.96	&	0.522	$\pm$	0.042	\\
53111.330	&	241.57	&	168.73	&	0.518	$\pm$	0.041	\\
\enddata
\end{deluxetable}

\clearpage
\begin{deluxetable}{clccccccccccccc}
\tabletypesize{\scriptsize}
\rotate
\tablewidth{0pc}
\tablecaption{Parameters of Stellar Models for Regulus\label{tab11}}
\tablehead{
\colhead{ }               &
\colhead{ }               &
\multispan{7}{\hfil From~Spectroscopy \hfil} &
\colhead{ }               &
\multispan{5}{\hfil From~Interferometry \hfil} \\
\cline{3-9}               
\cline{11-15}             
\colhead{$i$}             &
\colhead{ }               &
\colhead{$V \sin i$}      &
\colhead{ }               &
\colhead{$R_p$}           &
\colhead{$R_e$}           &
\colhead{$M$}             &
\colhead{$T_p$}           &
\colhead{$T_e$}           &
\colhead{ }               &
\colhead{$R_{\rm minor}$} &
\colhead{$R_{\rm major}$} &
\colhead{$\alpha$}        &
\colhead{}                &
\colhead{}                \\
\colhead{(deg)}           &
\colhead{$\beta$}         &
\colhead{(km s$^{-1}$)}   &
\colhead{$V_e/V_c$}       &
\colhead{($R_\odot$)}     &
\colhead{($R_\odot$)}     &
\colhead{($M_\odot$)}     &
\colhead{(K)}             &
\colhead{(K)}             &
\colhead{ }               &
\colhead{(mas)}           &
\colhead{(mas)}           &
\colhead{(deg)}           &
\colhead{$d/d(Hipparcos)$}&
\colhead{$\chi^2_\nu$}    }
\startdata
90  & 0.00  & 305 & 0.82 & 3.15 & 4.05 & 3.45 & 12280 &  12280 && 0.624 & 0.802 & 87.4 & 0.988 & 3.95 \\
90  & 0.05  & 307 & 0.82 & 3.15 & 4.06 & 3.45 & 12825 &  11968 && 0.624 & 0.806 & 87.2 & 0.987 & 3.76 \\
90  & 0.10  & 309 & 0.83 & 3.14 & 4.07 & 3.45 & 13400 &  11643 && 0.626 & 0.810 & 86.8 & 0.984 & 3.58 \\
90  & 0.15  & 310 & 0.83 & 3.14 & 4.08 & 3.46 & 14006 &  11309 && 0.625 & 0.811 & 86.1 & 0.984 & 3.45 \\
90  & 0.20  & 314 & 0.84 & 3.14 & 4.12 & 3.43 & 14660 &  10852 && 0.624 & 0.816 & 86.0 & 0.987 & 3.38 \\
90  & 0.25  & 317 & 0.86 & 3.14 & 4.16 & 3.39 & 15400 &  10314 && 0.623 & 0.825 & 85.5 & 0.988 & 3.35 \\
90  & 0.30  & 323 & 0.87 & 3.15 & 4.22 & 3.39 & 16235 &\phn9677&& 0.619 & 0.830 & 85.3 & 0.994 & 3.39 \\
90  & 0.35  & 326 & 0.88 & 3.14 & 4.24 & 3.39 & 17120 &\phn9155&& 0.620 & 0.836 & 84.6 & 0.993 & 3.55 \\
80  & 0.00  & 305 & 0.83 & 3.13 & 4.05 & 3.45 & 12280 &  12280 && 0.626 & 0.804 & 87.6 & 0.987 & 3.97 \\
80  & 0.05  & 307 & 0.83 & 3.12 & 4.06 & 3.45 & 12836 &  11943 && 0.627 & 0.807 & 87.2 & 0.985 & 3.78 \\
80  & 0.10  & 309 & 0.84 & 3.12 & 4.07 & 3.45 & 13420 &  11589 && 0.626 & 0.809 & 86.8 & 0.985 & 3.61 \\
80  & 0.15  & 310 & 0.84 & 3.12 & 4.08 & 3.45 & 14035 &  11220 && 0.628 & 0.814 & 86.6 & 0.982 & 3.48 \\
80  & 0.20  & 314 & 0.85 & 3.12 & 4.12 & 3.44 & 14686 &  10740 && 0.626 & 0.818 & 86.1 & 0.985 & 3.40 \\
80  & 0.25  & 317 & 0.87 & 3.12 & 4.16 & 3.41 & 15435 &  10191 && 0.625 & 0.825 & 85.9 & 0.987 & 3.36 \\
80  & 0.30  & 323 & 0.88 & 3.12 & 4.22 & 3.40 & 16280 &\phn9486&& 0.625 & 0.835 & 85.7 & 0.988 & 3.36 \\
75  & 0.00  & 305 & 0.84 & 3.10 & 4.05 & 3.46 & 12280 &  12280 && 0.626 & 0.803 & 87.7 & 0.988 & 3.98 \\
75  & 0.05  & 307 & 0.85 & 3.10 & 4.07 & 3.46 & 12845 &  11905 && 0.629 & 0.808 & 87.3 & 0.985 & 3.80 \\
75  & 0.10  & 309 & 0.85 & 3.10 & 4.08 & 3.46 & 13442 &  11516 && 0.629 & 0.810 & 86.9 & 0.985 & 3.64 \\
75  & 0.15  & 310 & 0.85 & 3.10 & 4.09 & 3.46 & 14068 &  11109 && 0.630 & 0.814 & 86.7 & 0.983 & 3.51 \\
75  & 0.20  & 314 & 0.86 & 3.10 & 4.12 & 3.44 & 14740 &  10581 && 0.630 & 0.821 & 86.5 & 0.983 & 3.43 \\
75  & 0.25  & 317 & 0.88 & 3.10 & 4.17 & 3.40 & 15510 &\phn9953&& 0.630 & 0.830 & 86.5 & 0.983 & 3.38 \\
70  & 0.00  & 305 & 0.86 & 3.06 & 4.06 & 3.46 & 12280 &  12280 && 0.628 & 0.804 & 87.6 & 0.987 & 4.02 \\
70  & 0.05  & 307 & 0.86 & 3.06 & 4.07 & 3.47 & 12862 &  11849 && 0.629 & 0.808 & 87.5 & 0.985 & 3.85 \\
70  & 0.10  & 309 & 0.87 & 3.06 & 4.08 & 3.46 & 13476 &  11401 && 0.630 & 0.813 & 87.3 & 0.982 & 3.70 \\
70  & 0.15  & 310 & 0.87 & 3.06 & 4.09 & 3.47 & 14119 &  10942 && 0.634 & 0.816 & 86.9 & 0.980 & 3.57 \\
70  & 0.20  & 314 & 0.88 & 3.06 & 4.13 & 3.45 & 14800 &  10336 && 0.635 & 0.825 & 87.1 & 0.979 & 3.50 \\
70  & 0.25  & 317 & 0.90 & 3.06 & 4.17 & 3.41 & 15585 &\phn9618&& 0.634 & 0.833 & 87.2 & 0.980 & 3.46 \\
\enddata
\end{deluxetable}

\clearpage
\begin{deluxetable}{lccc}
\tablewidth{0pc}
\tablecaption{Parameter Estimates from Different Models\label{tab4}}
\tablehead{
\colhead{}                  &
\colhead{Uniform}           &
\colhead{Grid}              &
\colhead{Spectroscopically} \\
\colhead{Parameter}         &
\colhead{Ellipsoid}         &
\colhead{Search}            &
\colhead{Constrained}       }
\startdata
No.\ parameters        \dotfill  & 3                & 6                &
4               \\ 
$R_{\rm minor}$ (mas)  \dotfill  & $0.651\pm0.016$  & $0.627\pm0.015$  &
$0.623\pm0.006$ \\
$R_{\rm major}$ (mas)  \dotfill  & $0.771\pm0.032$  & $0.810\pm0.023$  &
$0.825\pm0.008$ \\
$\alpha$ (deg)         \dotfill  & $84.9\pm2.4$\phn & $83.7\pm2.5$\phn &
$85.5\pm2.8$\phn\\
$i$ (deg)              \dotfill  & \nodata          & $85^{+5}_{-9}$   &
$90^{+0}_{-15}$ \\
$\beta$                \dotfill  & \nodata          & $0.13\pm0.05$    &
$0.25\pm0.11$   \\
$\chi^2_\nu$           \dotfill  & 3.41             & 3.73             &
3.35            \\
\enddata
\end{deluxetable}

\clearpage
\begin{deluxetable}{lc}
\tablewidth{0pc}
\tablecaption{Final Parameters for Regulus\label{tab5}}
\tablehead{
\colhead{Parameter}  &
\colhead{Value}      }
\startdata
$\Theta_{\rm minor}$ (mas)\dotfill &  $1.25\pm0.02$         \\
$\Theta_{\rm major}$ (mas)\dotfill &  $1.65\pm0.02$         \\
$\alpha$ (deg)            \dotfill &  $85.5\pm2.8$\phn      \\
$i$ (deg)                 \dotfill &  $90^{+0}_{-15}$       \\
$\beta$                   \dotfill &  $0.25\pm0.11$         \\
$V\sin i$ (km s$^{-1}$)   \dotfill &  $317\pm3$\phn\phn     \\
$V_e/V_c$                 \dotfill &  $0.86\pm0.03$         \\
$R_p$ ($R_\odot$)         \dotfill &  $3.14\pm0.06$         \\
$R_e$ ($R_\odot$)         \dotfill &  $4.16\pm0.08$         \\
$M$ ($M_\odot$)           \dotfill &  $3.4\pm0.2$           \\
$T_p$ (K)                 \dotfill &  $15400\pm1400$\phn    \\
$T_e$ (K)                 \dotfill &  $10314\pm1000$\phn    \\
$<T>$ (K)                 \dotfill &  $12901\pm500$\phn\phn \\
$L$ ($L_\odot$)           \dotfill &  $347\pm36$\phn        \\
$d$ (pc)                  \dotfill &  $23.5\pm0.4$\phn      \\
$A_V$ (mag)               \dotfill &  $0.016\pm0.010$       \\
\enddata
\end{deluxetable}


\end{document}